\documentclass[lettersize,journal]{IEEEtran}
\usepackage{amsmath,amsfonts}
\usepackage{algorithmic}
\usepackage{algorithm}
\usepackage{array}
\usepackage[caption=false,font=normalsize,labelfont=sf,textfont=sf]{subfig}
\usepackage{textcomp}
\usepackage{stfloats}
\usepackage{url}
\usepackage{verbatim}
\usepackage{graphicx}
\usepackage{cite}
\usepackage[switch]{lineno}
\usepackage{hyperref}
\hypersetup{
    colorlinks=true,
    linkcolor=blue,
    urlcolor=cyan,
    }

\newcommand\scalemath[2]{\scalebox{#1}{\mbox{\ensuremath{\displaystyle #2}}}}

\DeclareRobustCommand{\IEEEauthorrefmark}[1]{\smash{\textsuperscript{\footnotesize #1}}}

\begin{document}

\title{SigWavNet: Learning Multiresolution Signal Wavelet Network for Speech Emotion Recognition}

\author{\IEEEauthorblockN{Alaa Nfissi\IEEEauthorrefmark{1,2,4}\hspace{0.5cm}
Wassim Bouachir\IEEEauthorrefmark{1,4}\hspace{0.5cm}
Nizar Bouguila\IEEEauthorrefmark{2}\hspace{0.5cm}
Brian Mishara\IEEEauthorrefmark{3,4}}\\ \vspace{0.5em}
\IEEEauthorblockA{\textit{\IEEEauthorrefmark{1} Data Science Laboratory, University of Québec (TÉLUQ), Montréal, Canada}}\\
\IEEEauthorblockA{\textit{\IEEEauthorrefmark{2} Concordia Institute for Information Systems Engineering, Concordia University, Montréal, Canada}}\\
\IEEEauthorblockA{\textit{\IEEEauthorrefmark{3} Psychology Department, University of Québec at Montréal, Montréal, Canada}}\\
\IEEEauthorblockA{\textit{\IEEEauthorrefmark{4} Centre for Research and Intervention on Suicide, Ethical Issues and End-of-Life Practices, Montréal, Canada}% <-this % stops an unwanted space
}}

\maketitle

\begin{abstract}
In the field of human-computer interaction and psychological assessment, speech emotion recognition (SER) plays an important role in deciphering emotional states from speech signals. Despite advancements, challenges persist due to system complexity, feature distinctiveness issues, and noise interference. This paper introduces a new end-to-end (E2E) deep learning multi-resolution framework for SER, addressing these limitations by extracting meaningful representations directly from raw waveform speech signals. By leveraging the properties of the fast discrete wavelet transform (FDWT), including the cascade algorithm, conjugate quadrature filter, and coefficient denoising, our approach introduces a learnable model for both wavelet bases and denoising through deep learning techniques. The framework incorporates an activation function for learnable asymmetric hard thresholding of wavelet coefficients. Our approach exploits the capabilities of wavelets for effective localization in both time and frequency domains. We then combine one-dimensional dilated convolutional neural networks (1D dilated CNN) with a spatial attention layer and bidirectional gated recurrent units (Bi-GRU) with a temporal attention layer to efficiently capture the nuanced spatial and temporal characteristics of emotional features. By handling variable-length speech without segmentation and eliminating the need for pre or post-processing, the proposed model outperformed state-of-the-art methods on IEMOCAP and EMO-DB datasets. The source code of this paper is shared on the Github repository: \href{https://github.com/alaaNfissi/SigWavNet-Learning-Multiresolution-Signal-Wavelet-Network-for-Speech-Emotion-Recognition}{https://github.com/alaaNfissi/SigWavNet-Learning-Multiresolution-Signal-Wavelet-Network-for-Speech-Emotion-Recognition}.
\end{abstract}


\section{Introduction}
\IEEEPARstart{S}{peech} Emotion Recognition (SER) plays a pivotal role in human-computer interaction, fostering more sophisticated dialogues between machines and humans. Its implications span various domains, notably in emergency call centers, where SER systems evaluate an individual's stress or fear levels, thereby enhancing the accuracy of assessing the caller's condition \cite{nfissi2024unlocking}. This, in turn, facilitates more effective decision-making within call centers. Additionally, in healthcare, SER proves beneficial in identifying psychological disorders, potentially minimizing the risk of suicidal behaviors \cite{cdcp_2022}. Furthermore, the integration of SER into virtual AI chatbots opens avenues for providing personalized therapy through online interactions \cite{9544671}.

Speech represents an intricate and high-frequency signal encompassing information about the conveyed message, speaker characteristics, gender, language, and emotional content. However, challenges persist in discerning emotions from speech signals, primarily due to diverse speaking styles. Physiological studies underscore the significance of capturing the complete emotional trajectory, necessitating sufficiently long speech segments \cite{nugroho2022development}. In the initial stages, tasks centered on high-frequency signals typically involve extracting relevant time-frequency features before integrating them into machine learning (ML) algorithms, which are commonly tailored for fixed or limited input sizes \cite{el2011survey}. Managing speech signals in an emotional context poses a challenge of balancing high-frequency sampling with low-dimensional decision states (e.g., emotions). Consequently, many applications demand sophisticated approaches capable of extracting meaningful and concise information from speech signals to facilitate analysis and SER.

Numerous effective applications of ML in the context of speech utilize extracted features as inputs—a manually curated, condensed representation of the original signals. Frequently, these features encompass a spectrogram \cite{wang2020missing}, wavelet coefficient statistics \cite{li2019deep}, or other variations \cite{fink2020potential}. Despite their notable success, these frameworks require careful feature extraction, a potentially time-intensive task. Moreover, the extracted features may exhibit sensitivity to unforeseen noise or changing conditions, posing challenges in designing domain-invariant features. The establishment of such features, whether through post-processing or ML methodologies, remains an ongoing research inquiry \cite{michau2021unsupervised}.

In light of these feature extraction challenges, it's worth noting the excellent capabilities of Fourier and wavelet transforms for feature extraction and data dimensionality reduction, despite the intensive labor they require based on the dataset and problem at hand. The wavelet transform, with advantageous properties such as a linear time algorithm, perfect reconstruction, and customizable wavelet functions \cite{fahad2021survey}, emerges as a favorable choice for feature representation. However, its adoption in the machine learning community is limited, with methods like the Fourier transform and its variants being more commonly employed \cite{huang2022research}. This underutilization of the wavelet transform may stem from the challenge of designing and selecting appropriate wavelet functions, typically derived analytically using Fourier methods. Moreover, the array of available wavelet functions without a comprehensive understanding of wavelet theory can make the selection process daunting, prompting many to opt for simpler methods.

Building on this premise, the rise of convolutional neural networks (CNN) \cite{kiranyaz20211d} brought about a realization in early research that a temporal CNN could function as a digital filter, capable of learning convolution kernels similar to a Fourier transform or wavelets \cite{zhang2018deep}, and even acquiring sparse representations \cite{papyan2017convolutional}. Subsequently, there were proposals to confine the network to operations resembling the Fourier transform \cite{uteuliyeva2020fourier} or the wavelet transform, either with continuous wavelet transform \cite{li2021waveletkernelnet} or discrete wavelet transform \cite{liu2019multi} \cite{khalil2020end}. These studies collectively showcase that architectures or kernels inspired by spectral analysis yield superior results in supervised deep learning (DL) tasks. However, it's noteworthy that these approaches are rarely tailored for ML tasks, with the connection to spectral transformation often limited to the network architecture or the initialization of convolution kernels. Additionally, Fourier transform-based DL architectures face scalability challenges when dealing with larger input sizes.

We present SigWavNet, a new E2E DL model designed for SER that leverages the power of FDWT, cascade algorithms, conjugate quadrature filters, and coefficient denoising. Additionally, our model incorporates a suite of advanced techniques including 1D dilated CNN, spatial attention, Bi-GRU, and temporal attention. This paper is structured as follows. Section \ref{related_works} discusses related works, highlighting the advancements and limitations of existing methods. In Section \ref{proposed_method}, we introduce our proposed method, detailing its motivation, methodology, and the unique aspects that distinguish it from previous approaches. The experimental setup, datasets used, and evaluation metrics are elaborated in Section \ref{experiments_and_results}, followed by a presentation of our results and a comparative analysis with existing methods. An ablation study is conducted in Section \ref{ablation_study} to validate the significance of various components of our proposed model. Finally, Section \ref{conclusion} concludes the paper with a summary of our findings, contributions to the field, and potential avenues for future research.

\section{Related Works}
\label{related_works}

Within the domain of SER, researchers have strategically combined a variety of feature extraction methods and classification models to improve recognition efficacy. This includes the incorporation of handcrafted features into traditional ML models and DL models as classifiers. Moreover, a subset of approaches adopts an E2E framework where features are autonomously extracted from the raw waveform signal of speech. This methodological diversity represents a systematic effort in SER research to optimize recognition accuracy by harnessing a spectrum of feature extraction techniques and model architectures.

In \cite{gao2017speech}, a signal segmentation methodology was introduced, using the depth first search (DFS) algorithm to determine segment duration and overlap. Local features, including pitch, mel frequency cepstral coefficients (MFCC), line spectral pairs (LSP), intensity, and zero-crossing rate (ZCR), were extracted from each segment, followed by a process of smoothing and normalization. The authors also computed global features using the open-source media interpretation by large feature-space extraction (Open SMILE) toolkit \cite{eyben2016open}. For the classification task, they employed a linear kernel support vector machine (SVM) with sequential minimal optimization (SMO), demonstrating proficiency in classifying distinct emotions.

In a different approach, authors in \cite{zhang2017speech} presented an alternative method by advocating the use of 3-channel log-mel spectrograms as features for training a deep CNN. These channels incorporate static, delta, and delta-delta components, representing the first and second derivatives of the signal. This configuration draws an analogy to an RGB image, where the mel-spectrogram channels play a role similar to the red, green, and blue channels. Similarly, in \cite{zhao2019speech}, authors employed a combination of a local feature learning block (LFLB) and a long short-term memory (LSTM) model to extract features from both raw audio and log-mel spectrograms.

Numerous studies have embraced the integration of DL models to improve emotion classification. DL, grounded in the exploration of artificial neural networks, strives to amalgamate lower-level features into more abstract, high-level features, unveiling latent patterns within the data and thereby enhancing the classifier's recognition rate compared to the original features \cite{schuller2003hidden}. Some studies employ DL approaches like 1D CNNs to directly learn relevant features from raw speech. However, the deployment of DL networks raises concerns about their inherent instability and the substantial amount of data required for effective training \cite{rolnick2017deep}. In \cite{mustaqeem2019cnn}, researchers advocated for a deep-stride CNN customized for classifying emotions. This model transforms 1D audio signals into a 2D spectrogram through the Short-Time Fourier Transform (STFT), involving preprocessing steps such as adaptive threshold-based noise removal from the audio signals.

On the other hand, the scattering transform, introduced in works like \cite{mallat2010recursive} and \cite{mallat2012group}, functions as a deep convolutional network employing predefined kernels for convolution. This design imparts stability against temporal shifts and deformations in the feature representation of signals. Given the temporal spread of emotion cues in speech, the scattering transform is adept at robustly capturing temporal variations and cues. In \cite{anden2014deep}, researchers compute scattering coefficients in the log-frequency domain, ensuring frequency transposition invariance and fostering speaker-independent representations for speech recognition. The scattering transform's versatility extends to both 1D and 2D data processing. In \cite{anden2019joint}, a joint time-frequency scattering approach is introduced, incorporating multiscale frequency energy distribution into a time-invariant representation. Additionally, \cite{ghezaiel2021hybrid} combines two-layer scattering coefficients with CNN layers, providing a stable descriptor of speaker information extracted from raw speech.

Other methodologies employing wavelet analysis for scrutinizing speech signals have been introduced \cite{silva2009discriminative}. Wavelet analysis, grounded in a multi-resolution framework, aims to capture nonlinear interactions similar to vortex flows. This analytical technique finds applications in denoising, detection, compression, classification, and various other domains \cite{rao2018discrete} \cite{daubechies1990wavelet}. A notable application of wavelet analysis is evident in \cite{zao2014time}, where wavelet-based pH time-frequency vocal source features are extracted alongside MFCC and Teager-Energy-Operator (TEO) based features for emotion classification. Another instance is seen in \cite{muthusamy2015particle}, which extracts features such as PLP, MFCCs, linear predictive cepstral coefficients (LPCC), stationary wavelet transform features, wavelet packet energy, and entropy features for emotion recognition. In \cite{zheng2018effectiveness}, tuned Q-factor wavelet transform (TQWT) and wavelet packet transform (WPT) methods were utilized to predict the emotions of stroke patients.

Many studies mentioned here commonly segment input signals into fixed-length segments. This approach stems from the classification models' prerequisite for inputs of uniform dimensions. While this facilitates the extraction of prosodic and spectral features, it overlooks the concentration of emotion-related information in specific voice segments of audio signals \cite{cowie2001emotion}. Departing from this conventional approach, experiments in \cite{mansoorizadeh2007speech} demonstrated the superior performance of voiced-based (variable-length) segments over frame-based (fixed-length) segments for emotion recognition. Another foundational aspect of the proposed method is the adoption of wavelets. For discrete time-series signal analysis in dynamic signals like speech, both the discrete Fourier transform (DFT) and the discrete wavelet transform (DWT) are crucial. While DFT provides a frequency distribution, it falls short for dynamically changing signals. The STFT addresses this but lacks flexibility in window size. In contrast, DWT offers flexibility in window size based on frequency analysis and the freedom to select the analysis function, as highlighted in a study comparing DFT and DWT \cite{steinbuch2005wavelet}.

Several efforts dedicated to SER through wavelets have been documented. For instance, in \cite{palo2018wavelet} wavelet coefficients contribute to features such as LPCC and MFCC for emotion classification. Notably, \cite{abdel2020egyptian} incorporates wavelet features alongside prosodic and spectral features for classifying Arabic speech emotions, emphasizing the use of an autoencoder for dimensionality reduction. In another study \cite{wang2020wavelet}, researchers delve into the use of wavelet packet coefficients for SER, presenting a comparison with MFCC features. Their experiments, employing a sequential floating-forward search method for feature selection, underscore the superior performance of classifiers trained with wavelet features. Additionally, \cite{kishore2013emotion} incorporates sub-band cepstral (SBC) features derived from the WPT, reporting enhanced accuracy compared to MFCC features. Moreover, \cite{huang2019feature} introduces sub-band spectral centroid weighted wavelet packet cepstral coefficients for emotion classification, demonstrating the effectiveness of a deep belief network when combining these features with prosody and voice quality features, especially in noisy conditions. Notably, the utility of wavelet packets extends to real-world noise conditions, as evidenced in \cite{vasquez2015emotion}. Conclusively, the integration of wavelet analysis in diverse studies signifies its efficacy in capturing nuanced features within speech signals, contributing to advancements in emotion classification and related tasks.

\section{Proposed Method}
\label{proposed_method}
\subsection{Motivation}

Despite advancements in SER, persistent challenges such as system complexity, inadequate feature distinctiveness, and vulnerability to noise interference remain unaddressed. These challenges not only compromise the accuracy of emotion recognition but also limit the applicability of SER systems in real-world environments. A primary motivation for our study is the inherent complexity associated with emotional speech processing. Current systems often fail to effectively manage the intricacies of emotional expressions in speech, necessitating the development of more sophisticated and nuanced methodologies. Furthermore, the limited distinctiveness of features in conventional SER approaches presents another significant challenge. Traditional methods, predominantly based on fixed-length frame segmentation, struggle to capture the full spectrum of emotional cues dispersed throughout speech. This inadequacy underscores the importance of developing advanced feature extraction strategies that can accurately identify emotional nuances in continuous voice segments. Additionally, the susceptibility of existing SER frameworks to noise interference highlights a critical area for improvement. In practical scenarios, where speech signals are frequently exposed to varied and noisy conditions, enhancing the noise robustness of SER systems is essential for reliable performance. Informed by insights from physiological and psychological research, which suggest that emotional information in speech is not confined to specific segments but rather distributed across longer durations, our proposed E2E method diverges from conventional practices. It seeks to align more closely with the natural characteristics of emotional expression in speech, potentially improving the model's capability to capture and recognize emotions more effectively. This novel approach aims to pave the way for more accurate, robust, and versatile SERs.

\subsection{Preliminary concepts}

\subsubsection{ Fast discrete wavelet transform (FDWT)}
In our method, we have centered our focus on a particular linear time-frequency transform, namely the wavelet transform. In SER and signal processing, the Uncertainty Principle reveals inherent limitations in simultaneously capturing the temporal and frequency details of a signal. This principle is mathematically grounded in the assertion that the localization of a function \(f\) and its DFT, \(Df\), cannot both be narrowly confined. Formally, this limitation is expressed in eq. \eqref{eq1_1}:
\begin{equation}
\|f\|_0 \cdot \|Df\|_0 \geq n
\label{eq1_1}
\end{equation}

\noindent where \(\|f\|_0\) denotes the sparsity of \(f\), measuring the count of non-zero elements, and \(\|Df\|_0\) represents the sparsity of its DFT. This equation indicates that a function and its Fourier transform cannot simultaneously be sparse, underscoring a fundamental trade-off between time and frequency localization. Drawing from quantum mechanics, the Heisenberg Uncertainty Principle \cite{busch2007heisenberg} analogously articulates that the probability distributions of a particle's position and momentum—mutually Fourier pairs—cannot both be sharply localized. This phenomenon constrains our ability to precisely determine both attributes concurrently as formulated in eq. \eqref{eq1_2}:
\begin{equation}
\Delta x \cdot \Delta p \geq \frac{\hbar}{2}
\label{eq1_2}
\end{equation}

\noindent with \(\Delta x\) and \(\Delta p\) representing the standard deviations of position and momentum distributions, respectively, and \(\hbar\) being the reduced Planck's constant. This principle highlights the inherent trade-offs in measuring the complementary properties of a system.

In response to these challenges, the Wavelet Transformation emerges as a compelling solution within signal processing. Unlike the Fourier transform, wavelets provide a multi-scale decomposition of a signal that achieves localized analysis in both time (or space) and frequency. This duality is essential for effectively dissecting complex signals like speech, where emotional nuances are in both temporal and spectral dimensions. The mathematical foundation of the Wavelet Transformation is established on the principle of dilating and translating a mother wavelet \(\psi(t)\) to capture signal characteristics across different scales as shown in eq. \eqref{eq1_3}:
\begin{equation}
\psi_{a,b}(t) = \frac{1}{\sqrt{|a|}} \psi\left(\frac{t-b}{a}\right)
\label{eq1_3}
\end{equation}

\noindent where \(a\) and \(b\) denote the scale and translation parameters, respectively. This formulation allows for an adaptive focus on signal features, navigating the constraints imposed by the Uncertainty Principle to enrich the analysis of emotional content in speech signals.
\begin{figure}[t]
\centering
\scalemath{0.4}{
\centerline{\includegraphics[width=1\textwidth]{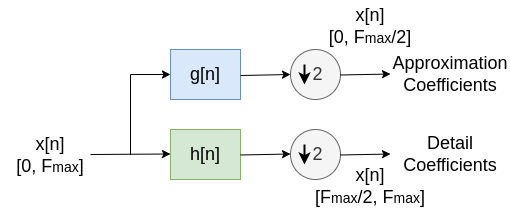}}
}
\caption{Traditional block diagram of wavelet filter analysis}
\label{fig1}
\end{figure}

Specifically, we utilize the FDWT, which employs a set of wavelets designed to create an orthonormal family through their scaling and translation by powers of 2, as delineated in \cite{mallat2008wavelet}. The core wavelet, or the 'mother wavelet' denoted as \(\psi\), is defined to have a zero mean and a unit norm. The variations of this wavelet function, through dilation and scaling, take the mathematical form in eq. \eqref{eq1}:
\begin{equation}
\psi_j[n]=\frac{1}{2^j} \psi\left(\frac{n}{2^j}\right)
\label{eq1}
\end{equation}

\noindent where, \(n, j \in Z\). The discrete wavelet transform for a given discrete real signal \(x\) is then defined in eq. \eqref{eq2} as:
\begin{equation}
W x\left[n, 2^j\right]=\sum_{m=0}^{N-1} x[m] \psi_j^*[m-n]
\label{eq2}
\end{equation}

In this context, wavelet functions serve as a band-pass filter bank, enabling the decomposition of a signal using this filter bank. As the wavelets are band-pass, we integrate a low-pass scaling function, \(\phi\), which cumulatively represents all wavelets above a certain scale, \(j\). The scaling function’s Fourier transform, \(\hat{\phi}\), adheres to the relationship in eq. \eqref{eq3}, with its phase being arbitrary \cite{mallat2008wavelet}:
\begin{equation}
|\hat{\phi}(\omega)|^2=\int_1^{+\infty} \frac{|\hat{\psi}(s \omega)|^2}{s} ds
\label{eq3}
\end{equation}

Utilizing these wavelets and the corresponding scaling function, the FDWT successively decomposes a signal \(x\) into a coarser approximation \(a\) (low-pass filtered version of \(x\)) and its detail coefficients \(d\) (high-pass filtered version of \(x\), also known as wavelet coefficients). With an orthonormal basis in \(\mathbf{L}^2(\mathbb{R})\), such a decomposition is reversible. Importantly, due to the scale factor of 2 between levels and in translating coefficients, the decomposition at any level \(l\) can be articulated as a function of the preceding approximation, sub-sampled by a factor of 2, and the base non-dilated wavelet.

The FDWT is computable via a fast-decimating algorithm, commonly referred to as the cascade algorithm. This is visually represented in the block diagram of a one-level FDWT cascade algorithm in Fig. \ref{fig1}, where \(g\) represents the wavelet and \(h\) the scaling function. We define two filters using eq. \eqref{eq4} and eq. \eqref{eq5}:
\begin{equation}
h[n]=\left\langle\frac{1}{\sqrt{2}} \phi\left(\frac{t}{2}\right), \phi(t-n)\right\rangle
\label{eq4}
\end{equation}
\begin{equation}
g[n]=\left\langle\frac{1}{\sqrt{2}} \psi\left(\frac{t}{2}\right), \phi(t-n)\right\rangle
\label{eq5}
\end{equation}

\noindent These equations establish a connection between the wavelet coefficients and the filters \(h\) and \(g\), leading to a recursive algorithm for computing the wavelet transform. This framework forms the foundation of our method in analyzing speech signals for emotion recognition.

\subsubsection{Conjugate quadrature filter (CQF)}
Our method leverages a notable feature of the FDWT related to filter identification, grounded in the principles of the CQF bank, as elucidated in \cite{croisier1976perfect}. A key aspect of the CQF bank is its quadrature property, which ensures a symmetric response from the decomposition filters relative to the cutoff frequency, thereby imparting an antialiasing characteristic to the system. This is achieved by designing the filters in such a way that the wavelet function \(g\) becomes the alternating flip of the scaling function \(h\). This relationship is denoted in eq. \eqref{eq6} as:
\begin{equation}
g[n] =(-1)^n \cdot h[-n]
\label{eq6}
\end{equation}

\noindent where, \(h[-n]\) represents the \(n^{th}\) coefficient of \(h\) in a reversed sequence.

The process of wavelet filter bank decomposition is integral to our method. It involves the computation of approximation and detail coefficients, \(a\) and \(d\) respectively, using the following eqs. \eqref{eq7} and \eqref{eq8}:
\begin{equation}
a_{j+1}[p]=\sum_{n=-\infty}^{+\infty} h[n-2p] a_j[n]
\label{eq7}
\end{equation}

\begin{equation}
d_{j+1}[p]=\sum_{n=-\infty}^{+\infty} g[n-2p] a_j[n]
\label{eq8}
\end{equation}

\noindent The detail coefficients \(d\) are essentially the wavelet coefficients as defined earlier. These coefficients are computed recursively at each scale, beginning with \(a_0\) initialized with the signal \(x\). During each step of the algorithm, the signal is divided into its high- and low-frequency components. This division is achieved by convolving the approximation coefficients with the scaling filter \(h\) and the wavelet filter \(g\). The low-frequency component, thus obtained, serves as the input for the subsequent step in the algorithm. It is important to note that both \(a_i\) and \(d_i\) coefficients undergo downsampling by a factor of two at each iteration, reducing the computational load.

A significant advantage of this algorithm is its requirement of only two filters (scaling and wavelet filters), as opposed to needing an entire filter bank. This efficiency is achieved without compromising the ability of the wavelet transform to effectively partition the signal into distinct frequency bands as defined by the wavelet functions. This methodical approach allows for a more precise and computationally efficient analysis of the signal's frequency components, which is essential for accurate SER.

Continuing our methodological discussion, a crucial application of wavelet analysis in our framework is signal denoising. The foundational assumption in wavelet-based denoising is that structured signals, when decomposed under an appropriate wavelet basis, typically result in a sparse decomposition. This means they activate specific wavelet coefficients at particular times and levels of decomposition. Conversely, noise, which is inherently unstructured, tends to activate wavelet filters at any level, but usually with lower amplitudes. Therefore, denoising typically involves applying a hard thresholding function to the wavelet coefficients. However, determining the optimal thresholding parameters has historically been a complex challenge and a subject of extensive research.

\subsection{Method}

\begin{figure}[t]
\centering
\centerline{\includegraphics[width=0.5\textwidth]{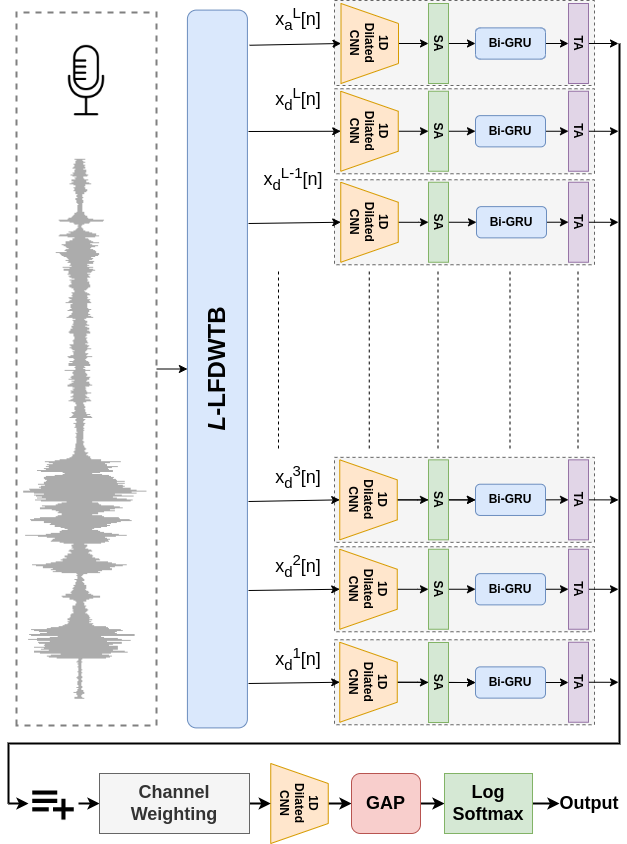}}
\caption{L-LFDWTB SigWavNet General Architecture}
\label{fig5}
\end{figure}

Our proposed architecture, SigWavNet, is an E2E deep learning framework tailored for SER, leveraging multiresolution analysis and advanced neural components to extract and interpret emotional features from speech signals. The foundation of the model, as shown in Fig. \ref{fig5}, is the learnable FDWT layer, which performs a multilevel decomposition of the raw speech waveform using pairs of learnable low-pass (\(Conv_h\)) and high-pass (\(Conv_g\)) filters. These filters are initialized as wavelet coefficients and refined during training for data-driven adaptation. A Learnable Asymmetric Hard Thresholding (LAHT) function is applied to the wavelet coefficients at each level, enabling dynamic denoising and enhancing the sparsity of the signal representation. This process emphasizes low-frequency variations critical for emotional content while isolating high-frequency details that capture transient features. The extracted multilevel features are then processed through dilated 1D dilated convolutional layers with spatial attention to capture local dependencies and prioritize emotionally significant regions. A Bidirectional Gated Recurrent Unit (Bi-GRU) network with temporal attention further refines the sequential patterns, emphasizing key temporal regions contributing to emotion recognition. Finally, a channel weighting mechanism combines multiband features, followed by a Global Average Pooling (GAP) layer and a Log Softmax layer to output emotion probabilities. By integrating data-driven multiresolution analysis, learnable thresholding, and attention mechanisms, SigWavNet effectively captures the complex and hierarchical structure of emotional cues in speech signals.

\begin{figure*}[t]
\centering
\centerline{\includegraphics[width=1\textwidth]{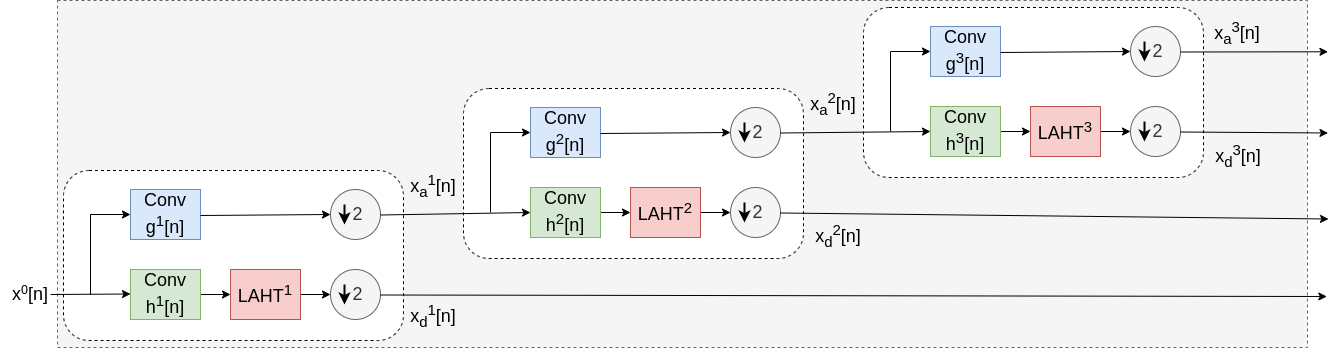}}
\caption{SigWavNet Three Levels Learnable Fast Discrete Wavelet Transform Block (3-LFDWTB)}
\label{fig2}
\end{figure*}

\subsubsection{Learnable FDWT}

In our implementation, we introduce a series of convolutional layers, \(Conv_h\) and \(Conv_g\), that function as learnable scaling (low-pass) and wavelet (high-pass) filters at each decomposition level. These layers facilitate the automatic extraction of meaningful and sparse representations from the raw speech waveform, tailored to the nuances of SER. To enhance the interpretability and effectiveness of the learnable FDWT, we initialize the \(Conv_h\) and \(Conv_g \) filters with a suitable wavelet, which provides an orthogonal wavelet basis designed for time-frequency analysis. This initialization helps to guide the network towards efficient signal decomposition while still allowing the filters to adapt optimally for the SER task during training.

\textbf{Convolutional Operations and Learning Principle:}
In traditional FDWT, fixed filters are used to compute the approximation and detail coefficients at each decomposition level. In our learnable FDWT framework, however, we implement these filters as convolutional layers \(Conv_h\) and \(Conv_g\), initialized with wavelet filter coefficients but further refined through training. For an input signal \( x[n] \), the low-pass (approximation) coefficients \( a_{j+1}[p] \) and high-pass (detail) coefficients \( d_{j+1}[p] \) at each level \( j \) are calculated as follows:
\begin{equation}
a_{j+1}[p] = \sum_{n=-\infty}^{+\infty} Conv_h(n - 2p) \cdot a_j[n]
\end{equation}
\begin{equation}
d_{j+1}[p] = \sum_{n=-\infty}^{+\infty} Conv_g(n - 2p) \cdot a_j[n]
\end{equation}

\noindent where:
\begin{itemize}
    \item \( a_j[n] \) represents the approximation (low-pass) coefficient from the previous level.
    \item \(Conv_h\) and \(Conv_g\) are initialized with low-pass and high-pass wavelet coefficients, such as Daubechies wavelets, but remain learnable for data-driven adaptation.
    \item The convolutional layers have a stride of 2, which performs downsampling as in traditional FDWT.
\end{itemize}

\textbf{Recursive Decomposition Process:}
The FDWT layer mirrors a recursive cascade structure across multiple levels \( L \), each consisting of \(Conv_h \) and \(Conv_g \) filters that adapt to capture SER-relevant features. At each level, \( a_{j+1} \) (low-pass result) and \( d_{j+1} \) (high-pass result) are obtained through the convolution operations, where \( a_{j+1} \) is passed to the next level for further decomposition. This recursive process captures frequency-specific features over time, resulting in a data-driven, multilevel wavelet decomposition optimized for the emotional features within the speech data.

\textbf{CQF Property:}  
To maintain orthogonality and ensure a coherent decomposition structure, we incorporate the CQF property between \(Conv_h\) and \(Conv_g\), simplifying the learning process while preserving wavelet theory principles. By deriving \(Conv_g\) from \(Conv_h\) as shown in eq. \eqref{eq6_1}, the number of parameters is halved, reducing optimization complexity and enhancing learning efficiency as proven in the ablation study section \ref{ablation_study}:  
\begin{equation}
Conv_g[n] = (-1)^n \cdot Conv_h[-n]
\label{eq6_1}
\end{equation}

\noindent The CQF property ensures orthogonality, time-frequency localization, and perfect signal reconstruction, which are crucial for retaining the subtle variations in speech signals critical to emotional context. Additionally, it acts as a form of regularization, limiting transformations to wavelet-like operations, thus improving generalization and interpretability. This parametrization allows the model to emphasize low-frequency emotional components while isolating high-frequency details, facilitating robust and structured representation learning.

\textbf{Data-Driven Adaptation to SER:}
The wavelet-initialized learnable FDWT filters provide a starting point, capturing basic time-frequency structure. However, during training, the filters adapt further to isolate frequency components particularly relevant for SER. By minimizing our supervised loss function, the model tunes \(Conv_h\) and \(Conv_g\) to emphasize emotional features, such as tone and pitch variations, that are integral to SER. Through this learning process, the FDWT layer achieves a \textit{multiband filter bank structure} that analyzes the speech signal across different frequency bands in a manner that aligns with the MEL scale. This resemblance arises because, like the MEL scale, our FDWT layer emphasizes lower frequency components crucial for emotional tones, while the high-pass filters isolate finer details within the emotional cues.

In particular, our learnable FDWT focuses on decomposing only the low-frequency representation of the previous level at each stage, giving progressively more emphasis to variations in the low-frequency range of the speech signal. This approach reflects the MEL scale’s emphasis on perceptually important lower frequencies, where emotional nuances often reside, while still retaining high-frequency components as details at each level. Each level of the FDWT layer thus serves as a learnable, data-adapted filter bank that provides a nuanced analysis across emotional frequency bands, capturing both the broad tonal structure in the low frequencies and the finer, higher-frequency details essential for robust SER.

\textbf{Comparison with Classical Convolution Layers:}
Learnable FDWT layers and classical convolution layers both use convolution operations but serve distinct purposes. FDWT layers act as a multiband filter bank inspired by wavelet theory, decomposing signals into frequency-specific components across multiple levels, with a focus on low-frequency variations crucial for emotional content. In contrast, classical convolution layers extract features across the entire signal without targeting specific frequency bands. FDWT layers adapt their filters to align with time-frequency localization, enabling detailed, frequency-sensitive analysis of emotional cues, unlike the broader feature extraction of classical convolutions.

\textbf{Efficient, Lightweight Architecture:}
Our design also benefits from the hierarchical nature of wavelet decomposition, enabling efficient representation with fewer parameters, by constructing an \(L\)-level cascade network with only \(2L\) convolutional filters (one pair of \(Conv_h\) and \(Conv_g\) per level). Since the limit for \(L\) is set by the nearest second logarithm of the smallest training input size, we achieve a lightweight architecture that balances complexity with performance. This approach results in a significantly smaller and more efficient model than traditional deep networks, which often rely on millions of parameters.

\begin{figure}[t]
\centering
\scalemath{0.5}{
\centerline{\includegraphics[width=1\textwidth]{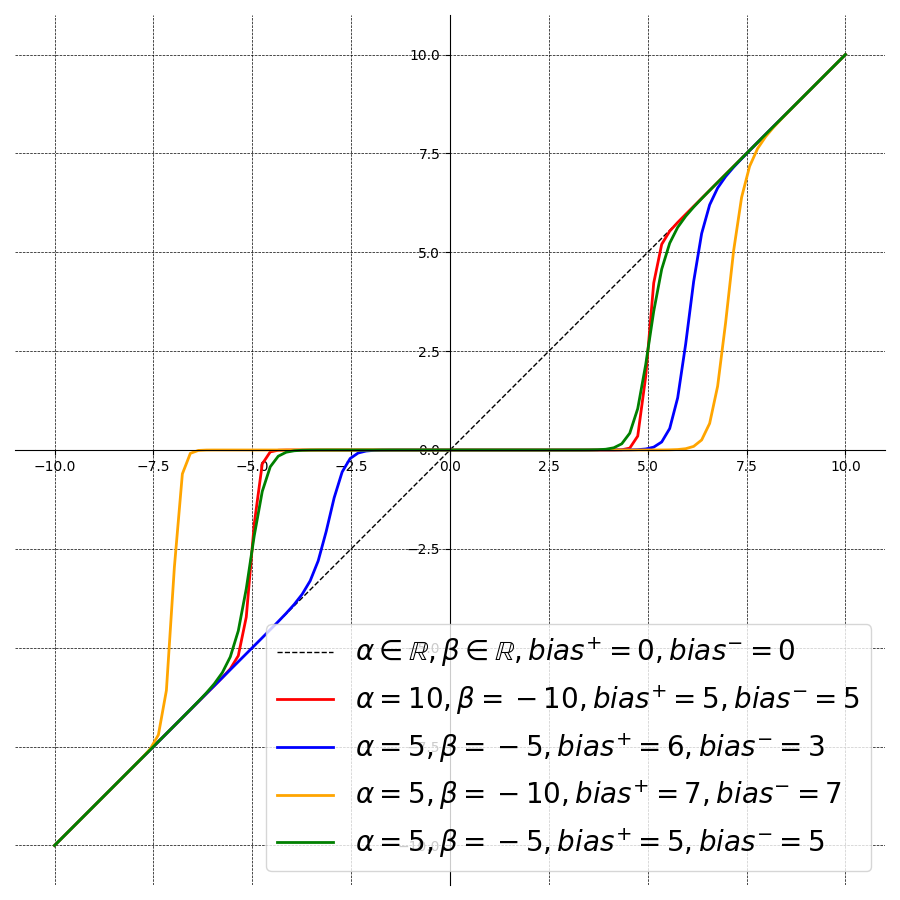}}
}
\caption{Learnable Asymmetric Hard Threshold (LAHT) Function}
\label{fig3}
\end{figure}

\subsubsection{Learnable asymmetric hard-thresholding (LAHT)}
To learn the correct hard-thresholding operation, the coefficients obtained are then processed by a specially designed LAHT layer/function. This layer is continuous and differentiable, mimicking a wavelet denoising operation, and feeds into the next block for further decomposition. This structure allows for a more nuanced and effective approach to signal denoising, enhancing the overall efficacy of the SER process.
\begin{figure*}[t]
\centering
\centerline{\includegraphics[width=1\textwidth]{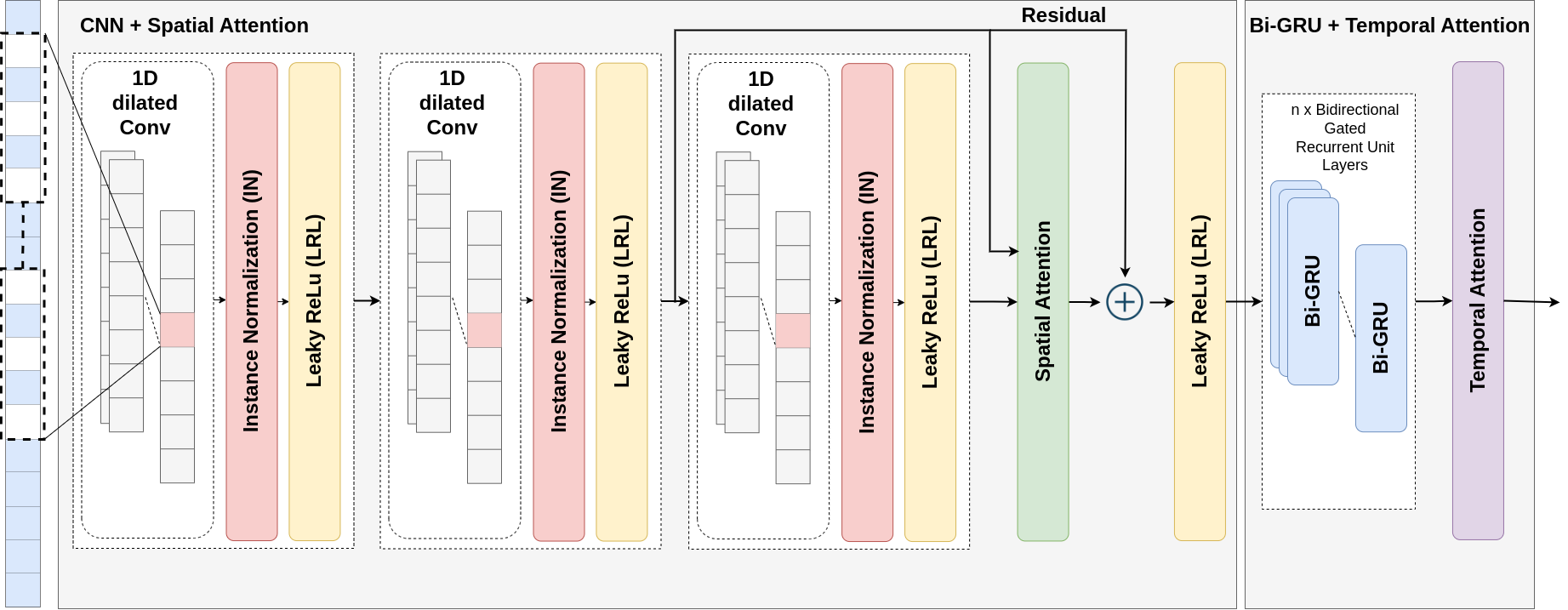}}
\caption{Architecture of a 1D Dilated CNN Enhanced with Spatial Attention, Followed by a Bi-GRU and Temporal Attention Mechanism. This diagram showcases the sequential processing flow from initial feature extraction through dilated convolutions, focus adjustment via spatial attention, sequential data handling with Bi-GRU, and critical temporal feature emphasis through temporal attention}
\label{fig4}
\end{figure*}

In our architecture, the thresholding step is integrated to autonomously learn the optimal thresholding parameters, eliminating the need for it to be treated as a separate process. We introduce the novel LAHT activation function, crafted as a blend of two contrasting sigmoid functions. This function is distinguished by its asymmetry in both sharpness and bias parameters. Mathematically, the sigmoid function \( S(x) \) is defined in eq. \eqref{eq9} as:
\begin{equation}
\forall x \in \mathbb{R}, \hspace{1cm} S(x)=\frac{1}{1+e^{-x}}
\label{eq9}
\end{equation}

\noindent Our LAHT function is then expressed in eq. \eqref{eq10} as:
\begin{equation}
\quad LAHT(x)=x \cdot\left[S\left(\alpha \cdot\left(x+bias^{-}\right)\right)+S\left(\beta \cdot\left(x-bias^{+}\right)\right)\right]
\label{eq10}
\end{equation}

\noindent In this equation, \( \alpha \) and \( \beta \) are the sharpness factors for negative and positive values, respectively, with the condition \( \alpha \cdot \beta < 0 \). The terms \( bias^{+} \) and \( bias^{-} \), both greater than 0, represent learnable biases that act as asymmetric thresholds on either side of the origin, as depicted in Fig. \ref{fig3}. To emulate the original FDWT without denoising, we can set both \( bias^{+} \) and \( bias^{-} \) to zero, while allowing \( \alpha \) and \( \beta \) to be real numbers, thus enforcing a linear activation.

This module's design features two principal distinctive elements: firstly, all convolution kernels, and secondly, both positive and negative hard thresholds, and sharpness factors are independent and asymmetrically learnable. Furthermore, our proposed sparsely connected deep neural network is capable of approximating representation systems that extend beyond the traditional wavelet representation system, offering greater generality and flexibility.

Each of the final low-pass and high-pass representations from all \(L\) decomposition levels is then forwarded to the subsequent module, ensuring a seamless and efficient flow of information within our DL framework. This integration of asymmetric learnability and efficient multi-resolution signal representation is pivotal to enhancing the capability and adaptability of our system in SER.

%%%%%%%%%%%%%%%%%%%%%%%%%%%%%%%%%%%%%%%%%%%%%%%%%%%%%%%
\subsubsection{1D dilated CNN with spatial attention and Bi-GRU with temporal attention}

The proposed architecture incorporates a 1D dilated CNN with a spatial attention mechanism and a Bi-GRU network with temporal attention to enhance feature extraction and sequence modeling for each frequency band, as shown in Fig. \ref{fig4}. The 1D dilated CNN efficiently captures local dependencies over varying time scales, while the spatial attention mechanism dynamically emphasizes the most relevant features across the input sequence. The Bi-GRU further models the sequential patterns by processing both forward and backward temporal dependencies, and the temporal attention layer identifies critical regions within the sequence that contribute most to emotion recognition. These components play a crucial role in complementing the proposed learnable FDWT layer. A detailed explanation of their design and implementation is provided in the Appendix.

\subsubsection{Channel weighting and global average pooling}
As we can see in the general architecture of SigWavNet illustrated in Fig. \ref{fig5}, the obtained attention vectors are subsequently concatenated channel-wise, to construct a composite representation with dimensions \((batch\_size, levels+1, length)\), where "\(levels+1\)" encompasses obtained frequency bands from high to low, and "\(length\)" denotes the dimensionality of each attention vector. Within this composite representation, each "\(channel\)" corresponds to an attention vector that stems from a specific frequency band of the original signal. This configuration coalesces the emotionally significant information harnessed from the obtained frequency spectrum of the speech signal, furnishing a rich, multifaceted feature set for the model's ensuing stages.

Following the concatenation of attention vectors into a composite representation with dimensions \((batch\_size, levels+1, length)\), our architecture incorporates a Channel Weighting layer, a novel component designed to further refine the aggregated feature set. This layer introduces a mechanism to dynamically adjust the importance of each channel (or frequency band) in the composite representation, allowing the model to emphasize or de-emphasize specific frequency components based on their relevance to emotion recognition as formulated in eq. \eqref{eq29}:

\begin{equation}
    X' = X \odot W
    \label{eq29}
\end{equation}

\noindent where \(X \in \mathbb{R}^{batch\_size \times (levels+1) \times length}\) denotes the input composite representation, \(W \in \mathbb{R}^{levels+1}\) symbolizes the learnable weights associated with each channel, and \(\odot\) signifies element-wise multiplication extended across the batch and length dimensions. The weights are optimized during the training process to learn the significance of each frequency band's contribution toward accurate emotion detection.

In this setup, `\textit{Channel Weighting}` defines a learnable parameter `\textit{weights}` for each channel, initially set to one, indicating equal importance across all channels. As the model undergoes training, these weights are adjusted to optimally scale the contribution of each channel based on how informative it is for the task at hand. This strategic weighting allows the model to leverage frequency-specific insights more effectively, enhancing its ability to discern and classify emotional states from speech signals with greater precision.

%%%%%%%%%%%%%%%%%%%%%%%%%%%%%%%%%%%%%%%%%%%%%%%%%%%%%%%%%%

Subsequently, another single 1D dilated convolutional layer is integrated, equipped with Instance Normalization (IN) and the Leaky ReLU activation function. This layer is crucial for its output channels, which match the number of classes (emotions) in our dataset. It serves to extract features from attention vectors' composite representation and generate feature maps corresponding to each emotion class. This design compels the network to learn distinct and accurate representations for each class in individual channels.

After the convolutional processing, the output is then directed to a Log Softmax layer for classification purposes. The Log Softmax layer, defined by eq. \eqref{eq30}, offers significant benefits over the traditional Softmax function:

\begin{equation}
\log \operatorname{Softmax}\left(x_i\right)=\log \left(\frac{\exp \left(x_i\right)}{\sum_j \exp \left(x_j\right)}\right)
\label{eq30}
\end{equation}

\noindent The superiority of Log Softmax lies in its enhanced numerical stability and computational efficiency. This function effectively addresses numerical instabilities that can arise from exponentiating large or small input values, which is a common challenge in DL applications. Furthermore, Log Softmax simplifies the gradient calculations during backpropagation. This simplification not only reduces computational complexity but also accelerates the learning process, making it a more efficient choice for training deep neural networks.

\begin{figure}[t]
\centering
\scalemath{0.15}{
\centerline{\includegraphics[width=1\textwidth, angle=-90]{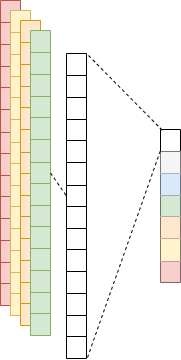}}
}
\caption{1D Global Average Pooling}
\label{fig6}
\end{figure}

To seamlessly integrate the final 1D dilated CNN layer and the Log Softmax layer, our architecture employs a single Global Average Pooling (GAP) layer, as advocated in \cite{lin2013network}. Unlike fully connected layers, the GAP layer averages the activations across the convolution output of the attention vectors. This process effectively condenses each feature map into a single scalar value, thereby reducing the dimensionality of the output while preserving essential information. The mechanism of the GAP layer is illustrated in Fig. \ref{fig6}.

The incorporation of GAP not only streamlines the network architecture by avoiding the complexity of fully connected layers but also contributes to reducing overfitting by minimizing the number of trainable parameters. This layer is particularly effective in summarizing the extracted features, ensuring that the subsequent Log Softmax layer receives a concise yet informative representation of the input for robust emotion classification.

\section{Experiments and Results}
\label{experiments_and_results}
\subsection{Datasets}

Datasets play an integral role in the development and effectiveness of SER systems. Generally, emotional speech databases are categorized into three types: actor-based, induced (also known as elicited or semi-natural), and natural emotional databases. The authenticity of these datasets is crucial for the model's ability to accurately recognize real-life emotions. Although natural datasets are most reflective of genuine emotional expressions, they often face copyright and security issues \cite{el2011survey}. These databases vary in terms of language, type and purpose, range of emotions, and demographics of speakers, including the number of speakers, samples, and utterances. A significant portion, about 66\%, of research-oriented databases use actor-based recordings. This trend is likely due to the controlled and diverse emotional representation these environments provide. In terms of language distribution, English is the most prevalent language in these databases, followed by Chinese and German. Most databases typically focus on fundamental emotions such as neutral, sad, happy, and angry, which are easier to simulate and recognize compared to more complex emotional states \cite{singh2022systematic}.

\subsubsection{\textbf{IEMOCAP}}
The Interactive Emotional Dyadic Motion Capture IEMOCAP dataset \cite{busso2008iemocap} offers a rich multi-modal and multi-speaker collection with 12 hours of audiovisual data, including video, voice, facial motion, and text transcriptions. It features dyadic sessions of improvised and scripted interactions designed to evoke various emotions, structured across five sessions with pairs of male and female speakers, totaling 10039 utterances each, averaging 4.5 seconds, at a 16 KHz sampling rate. Emotions in IEMOCAP are categorized by multiple annotators into basic emotions like anger, happiness, sadness, and neutrality, and dimensionally represented by valence, activation, and dominance scales. Notably, the dataset has class imbalances, with emotions like 'disgust', 'fear', and 'surprise' being less represented. Following prior research \cite{7178872} \cite{kim2013deep}, we exclude these classes and combine 'happy' and 'excited' into a single category to enhance dataset balance and research alignment.
\subsubsection{\textbf{EMO-DB}}
The Berlin Emotional Speech Database (EMO-DB) \cite{burkhardt2005database} is a pivotal SER resource, featuring 535 recordings in German by 10 professional actors (5 male and 5 female), simulating seven emotions: 'anger', 'boredom', 'disgust', 'anxiety/fear', 'happiness', 'sadness', and 'neutral'. Actors read 10 short, emotionally neutral texts in different emotional states to ensure recording consistency and avoid emotional bias. Each recording lasts approximately 5 seconds and was made in a professional, soundproof studio setting, ensuring high audio quality for analysis. Widely used in emotion recognition research, EMO-DB is freely available for non-commercial use, offering a valuable tool for SER technology development.

\subsection{Experimental setup}

To guarantee uniformity and ensure that our datasets are compatible, we first standardize all audio signals to a consistent format with a 16 KHz sampling rate and transform them into mono-channel. Following this standardization, we partition each dataset into two main subsets: 90\% allocated for training and validation, and the remaining 10\% reserved exclusively for testing as unseen data. To further refine the training and validation process, we adopted a 10-fold cross-validation approach. The segmentation of the data into these subsets, as well as the distribution within the cross-validation folds, was conducted using stratified random sampling \cite{aoyama1954study}. This technique divides the dataset into distinct, uniform groups—or strata—based on emotion classes, ensuring that each group is proportionately represented. This stratified approach contrasts with basic random sampling by ensuring that the selection of samples from each category is not merely random but proportionally representative, facilitating a division of the dataset that is both balanced and reflective.

To determine the best hyperparameters for our model, we employ a grid search technique. Hyperparameter optimization can be conducted through various approaches, one of which involves scheduling algorithms \cite{li2018massively}. These algorithms can manage trials by terminating problematic ones early, pausing, cloning, or adjusting the hyperparameters of ongoing trials. We chose the Asynchronous Successive Halving Algorithm (ASHA) due to its high efficiency and performance \cite{li2020system}.

In our pursuit to address the challenges presented by class imbalance and the propensity for model overfitting, we have adopted a loss function known as focal loss \cite{lin2017focal}, enhanced with L2-regularization. The focal loss function, originally designed to prioritize learning from hard-to-classify examples, is particularly effective in scenarios where the disparity between class distributions could otherwise skew the learning process. Its mathematical formulation is given by eq. \eqref{eq31}:
\begin{equation}
    FL(p_t) = -\alpha_t (1 - p_t)^\gamma \log(p_t)
    \label{eq31}
\end{equation}

\noindent where \(p_t\) is the model's estimated probability for a class, \(\alpha_t\) is a balancing factor, and \(\gamma\) is the focusing parameter that adjusts the rate at which easy examples contribute to the loss, thereby steering the model's attention towards more challenging instances.

To further refine our model's training dynamics and ensure its robustness against overfitting, we integrate L2-regularization (Ridge) into our loss function. This regularization technique imposes a penalty on the magnitude of the parameters, encouraging the model towards simpler, more generalizable patterns. The combined loss function, incorporating both focal loss and L2-regularization, is expressed in eq. \eqref{eq32}:
\begin{equation}
    J_{\text{regularized}}(\omega) = FL(p_t) + \lambda \sum_{i=1}^{n} \omega_i^2
    \label{eq32}
\end{equation}

\noindent where \(\lambda\) signifies the regularization strength, influencing the degree to which the model parameters are constrained.

By employing the Adam optimizer \cite{https://doi.org/10.48550/arxiv.1412.6980} to minimize this composite loss function over the training epochs, we ensure a balanced and effective optimization strategy. This amalgamation not only addresses the inherent issues posed by imbalanced datasets but also fortifies the model's capacity to generalize across a spectrum of emotional states in speech, thereby enhancing its predictive accuracy and reliability in real-world applications.

The initial weights of our model are set randomly, as it does not rely on any pre-trained models. We evaluate our model's performance using speaker-independent (SI) experiments on two publicly available datasets, IEMOCAP and EMO-DB. In the training phase, each sample is dedicated to predicting a single emotion. To ensure the accuracy of our findings, we repeat each test 10 times with different random seeds before calculating the average results.

\begin{figure}[t]
\centering
\scalemath{0.5}{
\centerline{\includegraphics[width=1\textwidth]{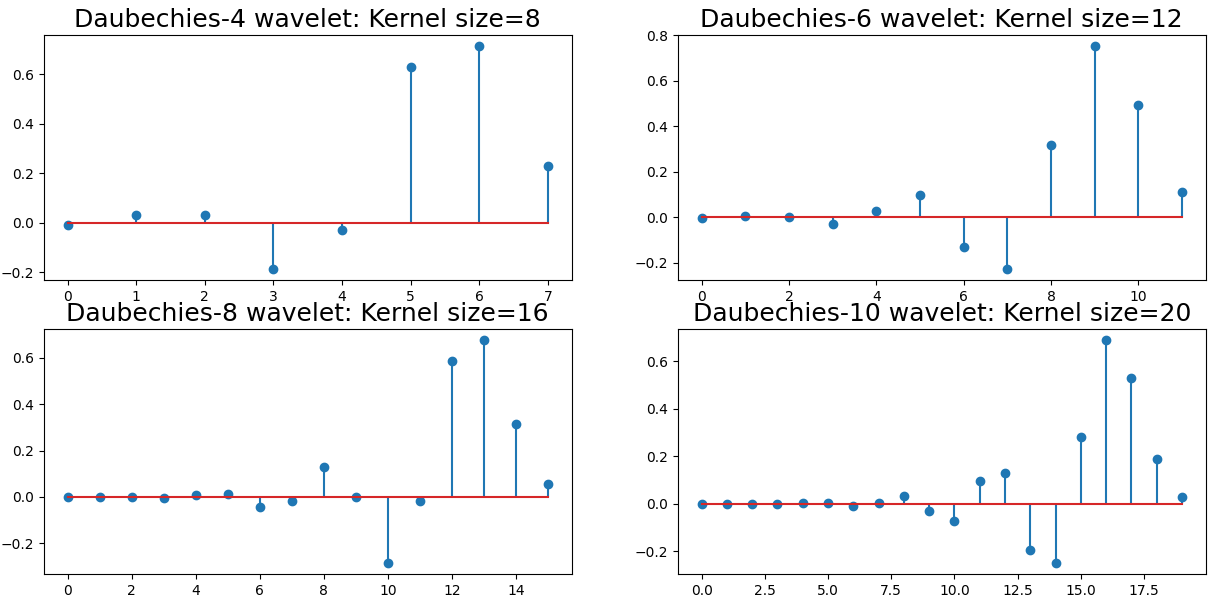}}
}
\caption{Daubechies Wavelets kernel size}
\label{fig7}
\end{figure}

As detailed in Section \ref{results}, the optimal architecture we arrived at includes an 8-LFDWTB with a kernel size of \(20\), emulating Daubechies-\(10\) wavelets (refer to Fig. \ref{fig7}) for the initial module. Daubechies orthogonal wavelets, ranging from D2–D20 or db1–db10, are commonly utilized. The index number indicates the number of coefficients, with each wavelet having a number of zero moments or vanishing moments equal to half of its coefficients (e.g., D8 has four vanishing moments, D20 has ten, and so on). The subsequent dilated CNN module consists of \(3\) convolutional layers, while the third module, which employs a Bi-GRU, incorporates \(6\) layers. This architectural configuration was determined based on the performance observed in our experiments. It represents the strategic selection of specific parameters for each module to enhance the model's efficacy in SER tasks. Notably, these parameters are flexible and can be tailored to suit various 1D signal classification tasks, allowing for customization to meet the unique requirements and characteristics of different signal processing applications beyond SER.

\subsection{Metrics}
In evaluating the performance of our model, we primarily focus on two key metrics: accuracy and F1-score. These metrics are essential for understanding how effectively our model predicts emotions in speech.

The accuracy metric is defined as the proportion of correct predictions made by the model out of the total number of predictions; it is formulated in eq. \eqref{eq33}:

\begin{equation}
    \textbf{\textit{Accuracy}} = \frac{\text{Number of Correct Predictions}}{\text{Total Number of Predictions}}
    \label{eq33}
\end{equation}

\noindent Accuracy is a straightforward and intuitive measure, providing a general indication of the model's performance across all classes. However, it may not always present a complete picture, especially in cases where the dataset is imbalanced.

The F1-score, on the other hand, offers a more nuanced evaluation. It is defined as the harmonic mean of precision and recall in eq. \eqref{eq34}:

\begin{equation}
    \textbf{\textit{F1-score}} = 2 * \frac{1}{\frac{1}{\text{precision}} + \frac{1}{\text{recall}}}
    \label{eq34}
\end{equation}

\noindent The F1-score balances the trade-off between precision and recall, providing a more holistic view of the model's performance, particularly in scenarios where the class distribution is uneven. Precision and recall are themselves defined as follows in eqs. \eqref{eq35} and \eqref{eq36}:

\begin{equation}
    \textbf{\textit{Precision}} = \frac{\text{True Positives}}{\text{True Positives} + \text{False Positives}}
    \label{eq35}
\end{equation}

\begin{equation}
    \textbf{\textit{Recall}} = \frac{\text{True Positives}}{\text{True Positives} + \text{False Negatives}}
    \label{eq36}
\end{equation}

Precision measures the proportion of actual positives among the instances the model predicted as positive, while recall quantifies how many actual positives the model successfully identified. By using both accuracy and F1-score, we ensure a comprehensive assessment of our model's performance, taking into account both its ability to correctly identify emotional states and its balance in handling different classes effectively.

\subsection{Results and comparison with state-of-the-art methods}
\label{results}

\subsubsection{\textbf{IEMOCAP}}

The data presented in Table \ref{table1} illustrates the effectiveness of the SigWavNet model in discerning various emotional expressions within the IEMOCAP dataset. This model demonstrates good accuracy in identifying different emotions, as reflected in its precision, recall, and F1-score metrics for each emotion category. In particular, SigWavNet excels in detecting 'Neutral' emotions with a high degree of precision (97\%) and recall (93\%), highlighting its capability to accurately identify this specific emotional state.

However, the model shows a noticeable precision-recall trade-off when it comes to emotions like 'Happy' and 'Sad'. For 'Happy', the model achieves a higher precision rate of 96.7\% at the cost of a lower recall. Conversely, in recognizing 'Sad' emotions, the model exhibits a higher recall rate of 95.4\% but with reduced precision. This variation indicates the model's differential sensitivity to various emotional expressions.

The macro and weighted average scores provide a holistic view of SigWavNet's overall performance. The macro average suggests a balanced performance across all emotion categories, while the weighted average takes into account the frequency of each class within the dataset. With an overall accuracy rate of 84.8\%, SigWavNet proves to be quite effective in the correct classification of emotions across the board. These findings underscore the robustness and potential applicability of SigWavNet in SER tasks, highlighting its capacity to distinguish and interpret a range of emotional states.

\begin{table}[t]
\centering
\caption{SigWavNet recognition performance over the IEMOCAP classes}
\begin{tabular}{|l|l|l|l|}
\hline
\textit{\textbf{Emotion}}  & \textbf{Precision (\%)} & \textbf{Recall (\%)} & \textbf{F1-score (\%)}   \\ \hline
\textbf{Angry}   &   65  &  80.9   &   72.1    \\ 
\textbf{Happy}   &   96.7   &  72   &   82.5     \\ 
\textbf{Neutral}  &   97      &  93    &   94.9      \\ 
\textbf{Sad}  &   79.2  &  95.4   &   86.6     \\ \hline

\multicolumn{4}{|c|}{\textbf{Overall results}}  \\ \hline
\textbf{Macro avg}   &  84.5   &  85.3   &   84      \\ 
\textbf{Weighted avg} &  87.1   &  84.8  &   85.1     \\ \hline
\textbf{Accuracy}          & \multicolumn{3}{c|}{84.8} \\ \hline
\end{tabular}
\label{table1}
\end{table}

\begin{figure}[t]
\centering
\scalemath{0.5}{
\centerline{\includegraphics[width=1\textwidth]{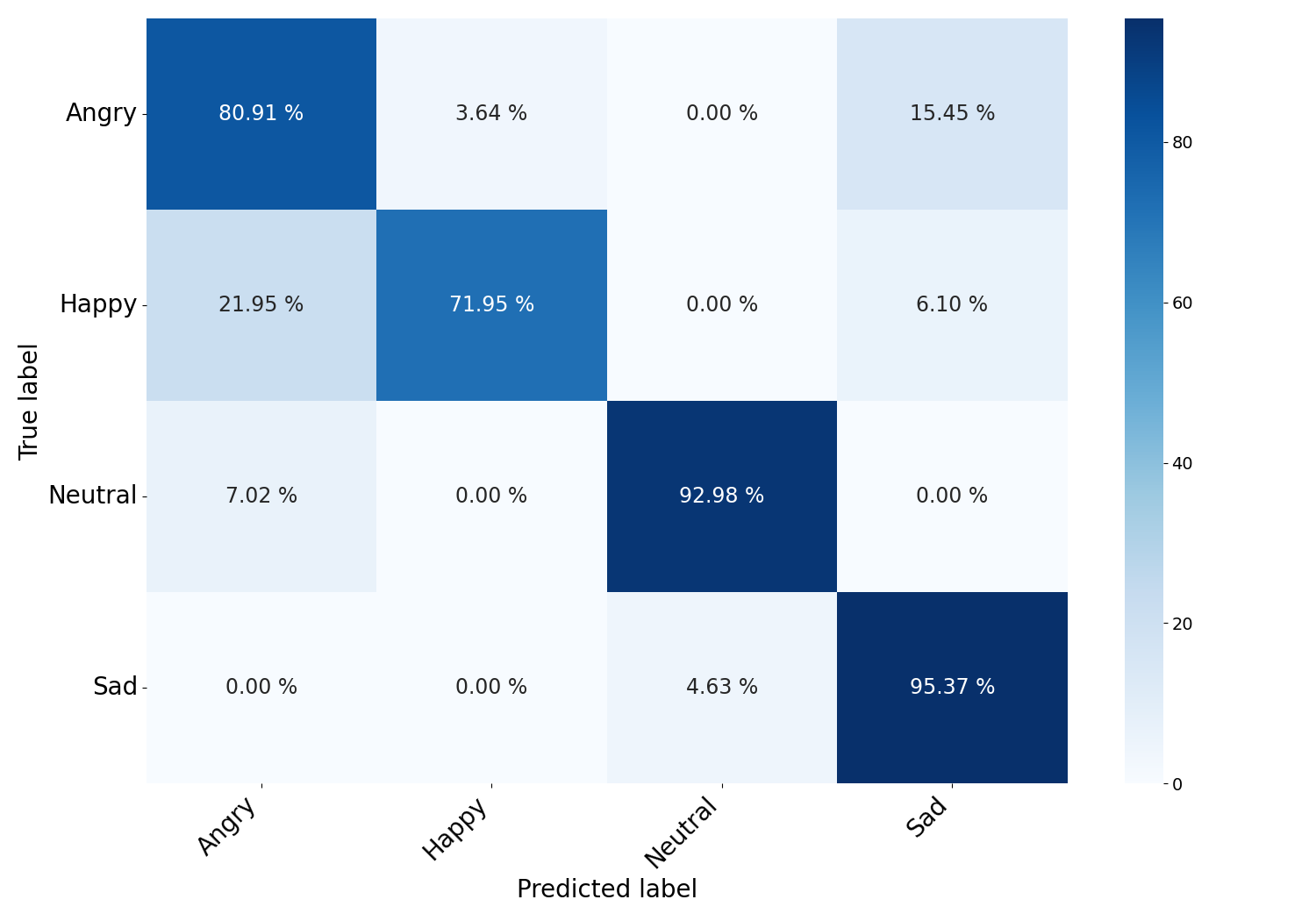}}
}
\caption{SigWavNet confusion matrix on IEMOCAP}
\label{fig8}
\end{figure}

Fig. \ref{fig8} presents a confusion matrix that offers an in-depth analysis of SigWavNet's capabilities in discerning various emotional states, revealing both its good accuracy and areas where enhancement is needed. In identifying 'Angry', the model achieves a good recognition rate of 80.91\%. However, it tends to confuse 'Angry' with 'Sad' in 15.45\% of cases, underscoring a challenge in differentiating these two emotions. Notably, there are no cases where 'Angry' is mistaken for 'Neutral', indicating a clear distinction the model makes between these two emotions. When it comes to recognizing 'Happy' emotions, SigWavNet successfully identifies them 71.95\% of the time. Yet, it incorrectly labels 'Happy' as 'Angry' in 21.95\% of instances, suggesting a difficulty in differentiating between these positive and negative emotional expressions. While the model is quite adept at detecting 'Happy' emotions, enhancing its recall could further improve its performance. SigWavNet excels in identifying 'Neutral' emotions, boasting a high correct recognition rate of 92.98\%. Its minimal misclassification with other emotional states demonstrates its strong capability of accurately distinguishing neutral expressions. The model also shows proficiency in recognizing 'Sad' emotions, with a correct recognition rate of 95.37\%. However, there are instances (4.63\%) where 'Sad' is misidentified as 'Neutral'. Despite this, the model's high accuracy rate suggests that its predictions for 'Sad' are generally reliable.

Overall, the confusion matrix in Fig. \ref{fig8} provides a comprehensive evaluation of SigWavNet's performance, emphasizing its strong suit in identifying 'Neutral' and 'Sad' emotions. While the model shows considerable effectiveness in certain emotional classifications, the observed misclassifications between 'Angry' and 'Sad', as well as 'Happy' and 'Angry', point to areas where the model could be refined. Enhancing the model's ability to distinguish these specific emotional nuances could greatly improve its overall effectiveness.

Table \ref{table2} details the performance of our model, SigWavNet, on the IEMOCAP dataset, showcasing its superior SI accuracy and F1-score in comparison to a variety of state-of-the-art methods in the field. These comparative methods utilize diverse techniques for feature extraction and classification. For instance, \cite{aftab2022light} applies MFCC for feature extraction. In contrast, \cite{issa2020speech} expands on this by incorporating MFCC along with four different spectral representations (Mel-scaled spectrogram, chromagram, spectral contrast feature, and Tonnetz representation) of the same record. The research by \cite{chen20183} focuses on using delta and delta-deltas for feature extraction, while \cite{meng2019speech} employs a 3-dimensional Log-Mel spectrum approach. When it comes to classification strategies, these methodologies predominantly rely on DL models. Both \cite{aftab2022light} and \cite{issa2020speech} utilize CNN, whereas \cite{li2019improved} integrates CNN with Bi-LSTM networks. The study by \cite{chen20183} introduces a 3-D attention-based convolutional RNN (ACRNN), merging CRNN with an attention mechanism. Meanwhile, \cite{meng2019speech} opts for dilated convolution networks as an alternative to traditional CNNs. Another approach is presented in \cite{singh2021deep}, where the scattering transform is used for SER, offering stable feature representations that are robust to deformations and shifts in time and frequency and adept at capturing emotional cues in speech. Additionally, \cite{meng2021speech} proposes a framework that combines multi-layer wavelet sequence sets derived from wavelet packet reconstruction (WPR) with standard feature sets, utilizing an RNN based on an attention mechanism for SER.

Our model, SigWavNet, stands out with the highest accuracy and F1-score among these state-of-the-art methods on the IEMOCAP dataset, as indicated in Table \ref{table2}. This performance is likely attributable to SigWavNet's capability to decompose the signal and amalgamate both local and temporally distributed features, resulting in a more effective representation of data for accurate SER.

\begin{table}[t]
\centering
\caption{Compared methods Performance on the IEMOCAP dataset: best results are in bold font}
\begin{tabular}{|l|l|l|}
\hline
\multicolumn{3}{|c|}{\textbf{Compared methods}}                                           \\ \hline
\textit{\textbf{IEMOCAP}}              & \textbf{Test Accuracy (\%)} & \textbf{F1-score (\%)} \\ \hline
\textbf{Iqbal et al. \cite{Iqbal2020MFCCAM}}         &   79                &  NA                \\ 
\textbf{Li et al. \cite{li2019improved}}   &    80.8    &  NA                \\ 
\textbf{Chen et al. \cite{chen20183}}       &    63.9           &  NA                \\ 
\textbf{Issa et al. \cite{issa2020speech}}       &    64.3           &  NA                \\ 
\textbf{Aftab et al. \cite{aftab2022light}}         &   70.78              &  78.84              \\
\textbf{Meng et al. \cite{meng2019speech}}       &    69.32           &  NA                \\ 

\textbf{Singh et al. \cite{singh2021deep}}       &    61.55          &  NA                \\ 
\textbf{Meng et al. \cite{meng2021speech}}      &    62.52         &  NA               \\ 
\textbf{SigWavNet (ours)}     &   \textbf{84.8}   &   \textbf{85.1}              \\ \hline
\end{tabular}
\label{table2}
\end{table}

\subsubsection{\textbf{EMO-DB}}

Table \ref{table3} offers a thorough assessment of SigWavNet's capacity to discern various emotions in the EMO-DB dataset, highlighting its good precision, recall, and F1-score metrics across different emotional states. For 'Anger', the model exhibits exceptional precision (100\%), signifying its high accuracy in predicting this emotion. Coupled with a recall of 92.3\%, it indicates SigWavNet's effectiveness in identifying most instances of 'Anger', resulting in a well-balanced F1-score of 96\%.

In recognizing 'Anxiety/Fear', SigWavNet achieves notable precision (94.7\%) and a slightly lower recall of 85.7\%, leading to a strong F1-score of 90\%. This demonstrates the model's aptitude for accurately identifying 'Anxiety/Fear' instances. The model also displays good accuracy in distinguishing 'Boredom', with precision and recall rates of 95.5\% and 87.5\%, respectively. The resulting F1-score of 91.3\% suggests that SigWavNet is adept at differentiating 'Boredom' from other emotional expressions. SigWavNet's recognition of 'Disgust' shows a balanced performance, with a precision of 73.7\% and a high recall of 93.3\%, resulting in an F1-score of 82.4\%. This indicates the model's efficiency in detecting 'Disgust' while minimizing false positives. For 'Happiness', the model performs well, achieving a precision of 90.9\% and a recall of 95.2\%, with an F1-score of 93\%, reflecting a balanced trade-off between precision and recall. In identifying 'Neutral' emotions, SigWavNet excels with a precision of 91.3\% and a recall of 87.5\%, leading to a balanced F1-score of 89.4\%. This shows the model's proficiency in accurately identifying neutral states. 'Sadness' recognition is also effective, with a precision of 76.2\% and a recall of 88.9\%, culminating in an F1-score of 82.1\%, which suggests a good balance between precision and recall, highlighting the model's capability in capturing 'Sadness' accurately.

When considering the macro average, SigWavNet demonstrates an overall balanced performance with average precision, recall, and F1-score values of 88.9\%, 90.1\%, and 89.2\%, respectively. The weighted average shows an overall high recognition performance with an accuracy of 90.1\%. Overall, the metrics presented in Table \ref{table3} affirm that SigWavNet is a robust model for emotion recognition across a diverse range of emotions in the EMO-DB dataset.

\begin{table}[t]
\centering
\caption{SigWavNet recognition performance over the EMO-DB classes}
\begin{tabular}{|l|l|l|l|}
\hline
\textit{\textbf{Emotion}}  & \textbf{Precision (\%)} & \textbf{Recall (\%)} & \textbf{F1-score (\%)}   \\ \hline
\textbf{Anger}             &   100            &  92.3         &   96    \\ 
\textbf{Anxiety/Fear}           &   94.7           &  85.7          &   90     \\ 
\textbf{Boredom}          &   95.5      &  87.5    &   91.3      \\ 

\textbf{Disgust}             &   73.7            &  93.3         &   82.4     \\ 
\textbf{Happiness}             &   90.9            &  95.2         &   93     \\ 
\textbf{Neutral}             &   91.3            &  87.5         &   89.4     \\ 
\textbf{Sadness}             &   76.2            &  88.9         &   82.1     \\ \hline

\multicolumn{4}{|c|}{\textbf{Overall results}}  \\ \hline
\textbf{Macro avg}         &  88.9            &  90.1         &   89.2      \\ 
\textbf{Weighted avg}      &  91.1            &  90.1         &   90.3      \\ \hline
\textbf{Accuracy}          & \multicolumn{3}{c|}{90.1} \\ \hline
\end{tabular}
\label{table3}
\end{table}

\begin{figure}[t]
\centering
\scalemath{0.5}{
\centerline{\includegraphics[width=1\textwidth]{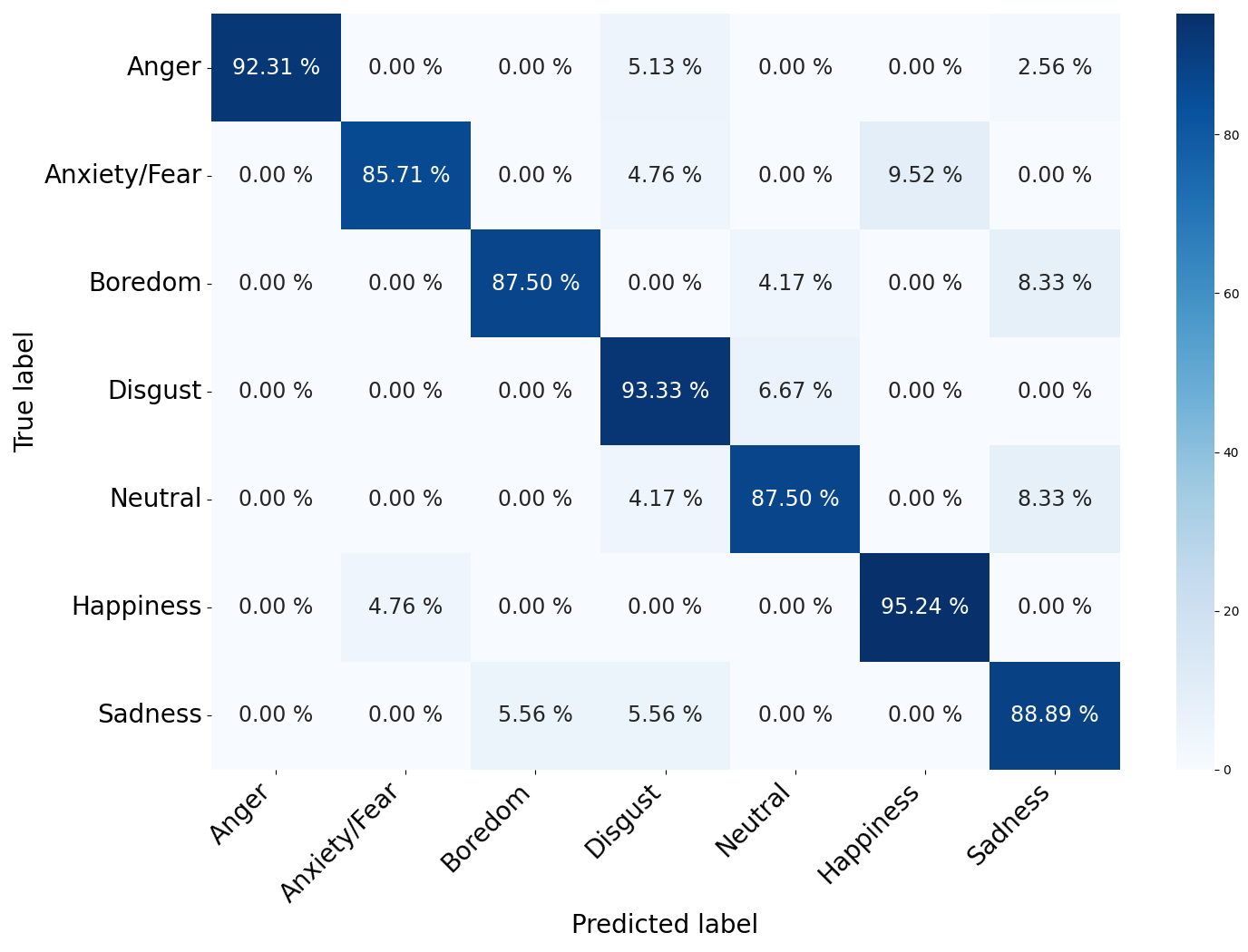}}
}
\caption{SigWavNet confusion matrix on EMO-DB}
\label{fig9}
\end{figure}

Fig. \ref{fig9} presents a confusion matrix that details SigWavNet's ability to identify various emotions in the EMO-DB dataset. SigWavNet shows good performance in detecting 'Anger', with a high correct recognition rate of 92.31\%. The model has minimal errors in this category, misidentifying 'Anger' as 'Disgust' in only 5.13\% of instances. When it comes to recognizing 'Anxiety/Fear', the model achieves a success rate of 85.71\%. However, there are instances where 'Anxiety/Fear' is confused with 'Happiness' (9.52\%) and 'Disgust' (4.76\%).

In identifying 'Boredom', SigWavNet maintains a solid performance, correctly recognizing this emotion in 87.50\% of cases. It occasionally misclassifies 'Boredom' as 'Neutral' (4.17\%) and 'Sadness' (8.33\%). The model's proficiency in detecting 'Disgust' is evident, with a high correct recognition rate of 93.33\%. There is a small tendency to confuse 'Disgust' with 'Neutral' (6.67\%), but no significant errors with other emotion classes, indicating a robust recognition ability.

SigWavNet also performs effectively in identifying 'Neutral' emotions, attaining a correct recognition rate of 87.50\%. Some instances of 'Neutral' are misinterpreted as 'Disgust' (4.17\%) and 'Sadness' (8.33\%). The model performs well at recognizing 'Happiness', with a high correct recognition rate of 95.24\%. It shows specificity in this classification, with a minor misclassification rate of 'Happiness' as 'Anxiety/Fear' (4.76\%) and no errors with other emotions. Regarding the recognition of 'Sadness', SigWavNet achieves a correct recognition rate of 88.89\%. Misclassifications occur with 'Sadness' being identified as 'Boredom' (5.56\%) and 'Disgust' (5.56\%).

Overall, the confusion matrix in Fig. \ref{fig9} offers a comprehensive view of SigWavNet's performance across different emotional states in the EMO-DB dataset. The model exhibits a high correct recognition rate in most classes, particularly in 'Anger', 'Disgust', 'Happiness', and 'Sadness'. The instances of misclassification provide insightful data, suggesting areas where the model can be further refined.

\begin{table}[t]
    \centering
    \caption{Compared methods Performance on the EMO-DB dataset: best results are in bold font}
    \begin{tabular}{|l|l|l|}
        \hline
        \multicolumn{3}{|c|}{\textbf{Compared methods}} \\ \hline
        \textit{\textbf{EMO-DB dataset}}              & \textbf{Test Accuracy (\%)} & \textbf{F1-score (\%)} \\ \hline
        \textbf{Parlak et al. \cite{parlak2014cross}} & 87.2  &  NA \\ 
        \textbf{Pham et al. \cite{pham2021emotion}} &    76.4 &  NA    \\ 
        \textbf{Ancilin et al. \cite{ancilin2021improved}} & 81.5 &  NA \\ 

        \textbf{Van et al. \cite{van2023speech}} & 71 &  NA \\ 
        
        \textbf{Tuncer et al. \cite{tuncer2021automated}} & 88.35 &  88.35 \\ 
        
        \textbf{Singh et al. \cite{singh2021deep}} & 74.59 &  NA \\ 
        
        \textbf{Palo et al. \cite{palo2023amalgamation}} & 77.4 &  NA \\  
        
        \textbf{Liu et al. \cite{liu2022speech}} & 89.13 &  89.4 \\ 

        \textbf{Meng et al. \cite{meng2021speech}} & 66.9 &  NA \\
        
        \textbf{SigWavNet (ours)} &  \textbf{90.1}  &  \textbf{90.3} \\ \hline
        \end{tabular}
      \label{table4}
\end{table}

In our analysis, we compare SigWavNet's SI performance with a selection of recent and significant studies. These studies represent a range of approaches and methodologies. The first study, \cite{parlak2014cross}, introduces EmoSTAR, a new emotional database, and conducts cross-corpus testing with the EMO-DB database. Utilizing the openSMILE toolkit for feature extraction and the WEKA tool for classification and feature selection, they achieve a test accuracy of 87.2\% on EMO-DB. The second study, \cite{pham2021emotion}, explores the use of DL for SER, particularly employing CNNs on the EMO-DB dataset. By employing various spectral features for acoustic signal analysis, they report an average accuracy of 99.3\% for binary classification and 76.4\% for the seven-class EMO-DB dataset. The third method, detailed in \cite{ancilin2021improved}, proposes an enhanced SER technique using Mel frequency magnitude coefficients for feature extraction. Testing their method across multiple databases, including EMO-DB, they report a notable accuracy of 81.5\% using a multiclass SVM classifier. Additional approaches include \cite{van2023speech}, which uses the fast Continuous Wavelet Transform (fCWT) to augment Deep CNN (DCNN) for SER. The study in \cite{palo2023amalgamation} presents a three-stage feature selection algorithm combining wavelet packet decomposition, statistical analysis, and an information gain (IG) ranking algorithm based on high IG entropy. In \cite{tuncer2021automated}, a novel nonlinear multi-level feature generation model with a cryptographic shuffle box structure is employed. The approach in \cite{liu2022speech} combines wavelet threshold denoising with a multi-task learning framework, utilizing shared and private LSTM networks for SER.

These methods are summarized in Table \ref{table4}, where we compare their SI performances on the EMO-DB dataset, highlighting the best results in bold. As per these comparisons, our proposed method, SigWavNet, surpasses these approaches, achieving an accuracy of 90.1\% and an F1-score of 90.3\%, demonstrating its superior performance in accurately recognizing emotions in speech.

\section{Ablation Study}
\label{ablation_study}
In this section, we conduct an ablation study to systematically examine the impact and contribution of various elements within our SigWavNet model for SER. This study involves a sequence of well-structured experiments, each aimed at understanding how different configurations affect the model's performance. These configurations include the use of Daubechies-10 wavelets without adaptable kernels, the implementation of learnable asymmetric hard thresholding, and varying approaches to kernel learning within the SigWavNet framework. By isolating and assessing these individual components, we aim to gain a deeper insight into how each contributes to the model's overall effectiveness in SER. This detailed exploration helps us identify the key factors that drive the success of our proposed approach, offering a clearer picture of the optimal model architecture.

\begin{figure*}[t]
\centering
\scalemath{1}{
\centerline{\includegraphics[width=1\textwidth]{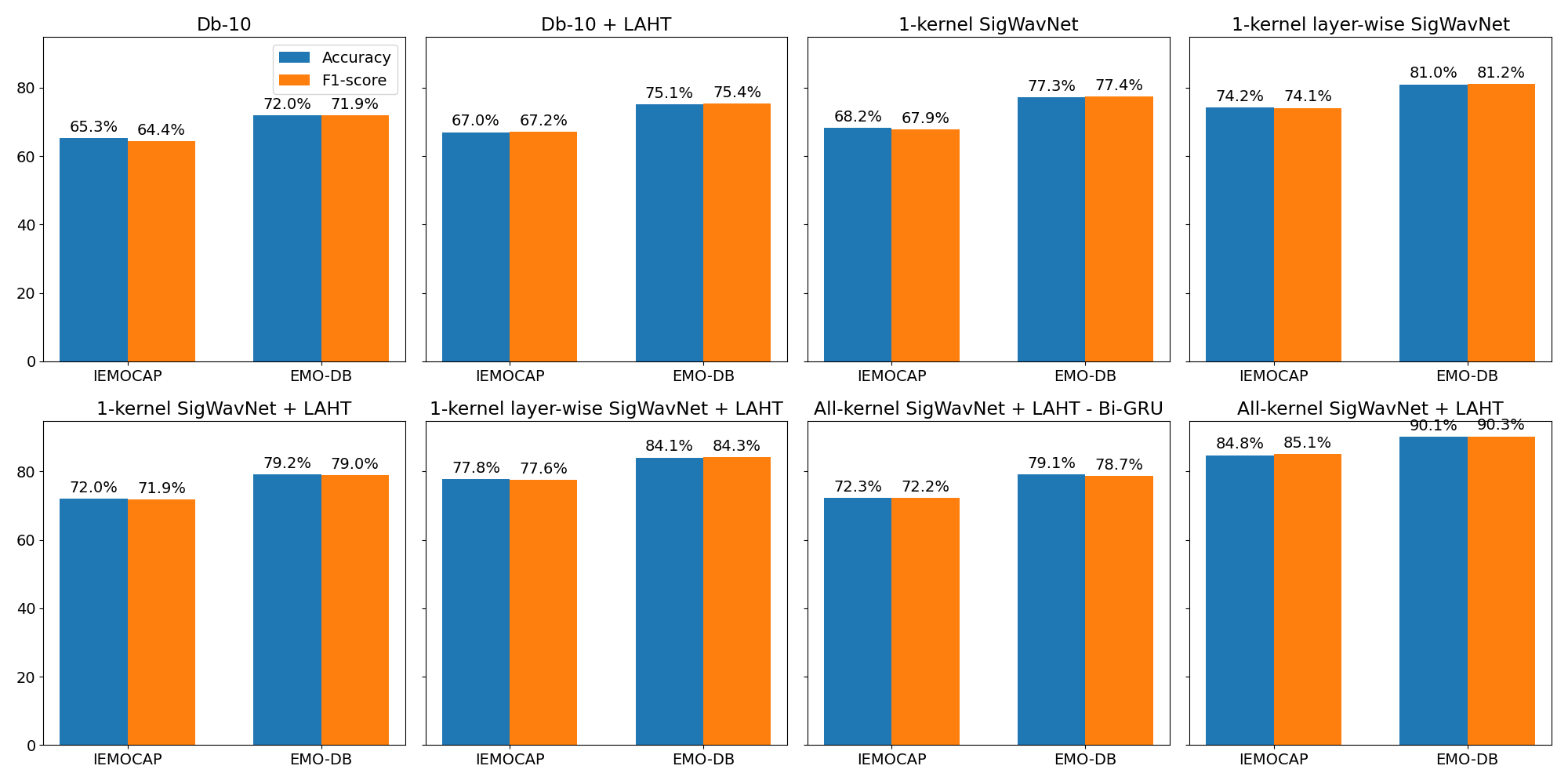}}
}
\caption{Ablation study configurations and performance}
\label{fig10}
\end{figure*}

The ablation study includes the following experiments:

\begin{itemize}
    \item \textbf{Db-10 (Non-Learnable Kernel):} This baseline experiment employs Daubechies-10 wavelets without any learnable kernels. It serves as a foundational comparison point for other configurations.

    \item \textbf{Db-10 + LAHT:} Building upon the baseline, this setup integrates learnable asymmetric hard thresholding. This addition aims to improve the adaptability of the model by effectively managing the thresholding of coefficients.

    \item \textbf{1-Kernel SigWavNet:} In this configuration, only the first kernel (\(h_0\)) is learned. The remaining kernels for various levels are deduced using the CQF technique.

    \item \textbf{1-Kernel Layer-Wise SigWavNet:} This variation involves learning a single kernel (\(h_l\)) for each level. The corresponding \(g_l\) for each layer is inferred using the CQF method, implementing a layer-wise approach.

    \item \textbf{1-Kernel SigWavNet + LAHT:} This model combines the single kernel learning strategy with learnable asymmetric hard thresholding. The aim is to enhance coefficient adaptation while maintaining a focused kernel learning.

    \item \textbf{1-Kernel Layer-Wise SigWavNet + LAHT:} This approach merges layer-wise single kernel learning with learnable asymmetric hard thresholding, aiming to boost the model’s adaptability and effectiveness.

    \item \textbf{All-Kernel SigWavNet + LAHT (Adopted Version):} In this configuration, all kernels are learned independently across different levels, and learnable asymmetric hard thresholding is integrated for superior adaptability.

    \item \textbf{All-Kernel SigWavNet + LAHT - (Bi-GRU \& Temporal Attention):} This configuration learns all kernels independently and incorporates learnable asymmetric hard thresholding but removes the Bi-GRU component and temporal attention.
\end{itemize}

Through this ablation study, we aim to dissect the contributions of each component in the SigWavNet architecture, shedding light on their individual and combined influence on the overall SER system, as demonstrated in Fig. \ref{fig10}. This in-depth analysis is expected to guide future optimizations.

\begin{figure}[t]
\centering
\scalemath{0.5}{
\centerline{\includegraphics[width=1\textwidth]{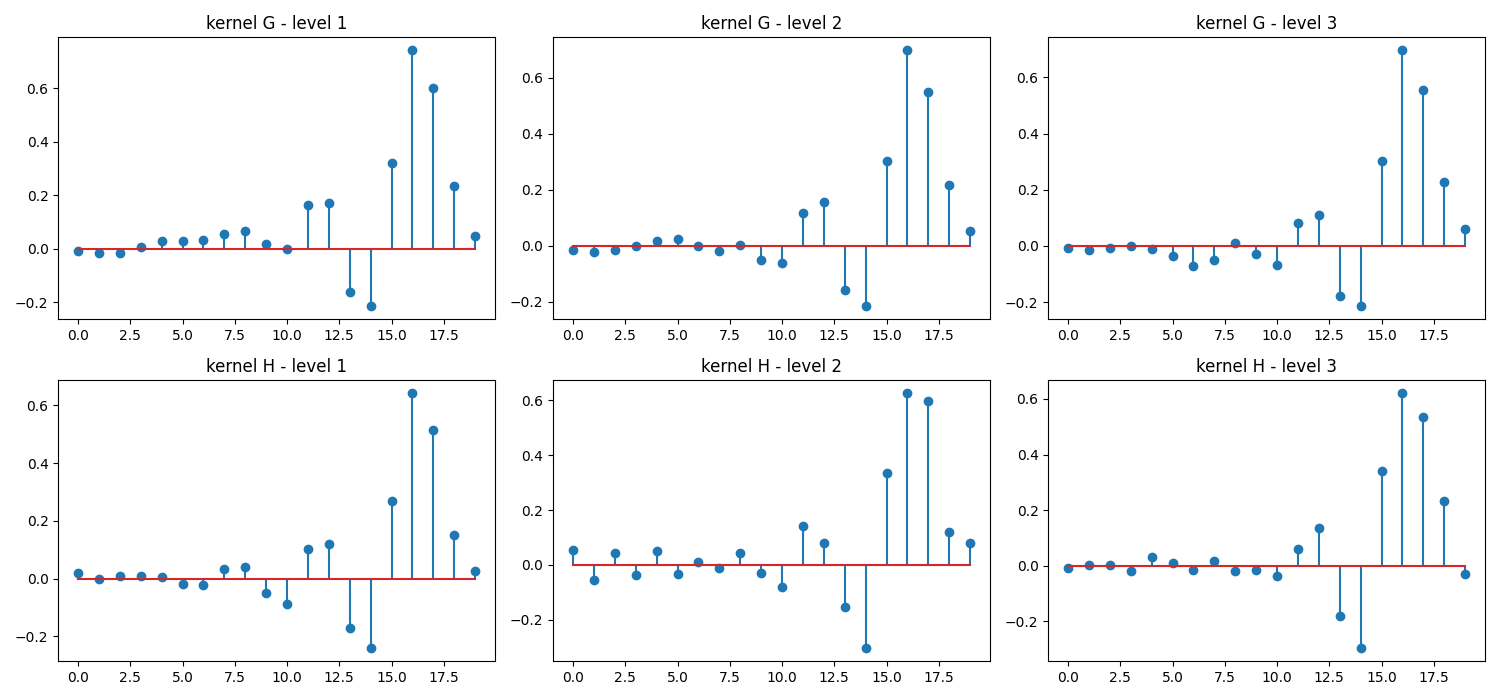}}
}
\caption{Low- and High-pass kernels of the first 3 levels after training}
\label{fig11}
\end{figure}

We begin with a baseline using Daubechies-10 wavelets without learnable kernels, achieving moderate accuracies of 65.3\% on IEMOCAP and 72\% on EMO-DB. This setup, lacking adaptability, provides a reference for evaluating enhancements. Incorporating LAHT improves adaptability, boosting accuracies to 67\% on IEMOCAP and 75.1\% on EMO-DB, highlighting the significance of dynamic thresholding in nuanced speech feature extraction. Exploring kernel learning strategies, the 1-kernel SigWavNet, which learns only the first kernel (\(h_0\)), shows the effectiveness of the CQF method, with accuracies of 68.2\% on IEMOCAP and 77.3\% on EMO-DB. The layer-wise approach of 1-kernel layer-wise SigWavNet further enhances performance, achieving 74.2\% on IEMOCAP and 81\% on EMO-DB. Integrating LAHT with single-kernel learning approaches yields further improvements. The 1-kernel SigWavNet + LAHT and the layer-wise variant both showed increased accuracies, underlining the effectiveness of this combination. The removal of the Bi-GRU and temporal attention in the All-kernel SigWavNet + LAHT - (Bi-GRU \& temporal attention) configuration leads to a performance decrease, underscoring the Bi-GRU and temporal attention role in capturing temporal dynamics and context. The most notable configuration is the All-kernel SigWavNet + LAHT, learning all kernels independently and incorporating LAHT. This configuration, which we adopt, achieves the highest accuracies, 84.8\% on IEMOCAP and 90.1\% on EMO-DB, demonstrating the synergy between comprehensive kernel learning and adaptive thresholding in effectively capturing diverse emotional nuances in speech signals as it takes full advantage of DL capabilities. In Fig. \ref{fig11}, we can see the kernels of the first 3 levels of signal decomposition after training, where we can see how they adapted to better extract relevant features for SER.

In conclusion, the ablation study offered valuable insights into each component's role, particularly highlighting the importance of LAHT, strategic kernel learning, and temporal aggregation in enhancing adaptability and capturing intricate speech patterns, leading to superior SER performance.

\section{Conclusion}
\label{conclusion}
In this work, we have presented SigWavNet, a novel architecture for Speech Emotion Recognition (SER), and evaluated its performance through an extensive series of experiments and ablation studies. Our findings have demonstrated that SigWavNet offers significant improvements in accuracy and F1-scores over existing methods, making it a robust and effective tool for SER. SigWavNet's architecture, which incorporates Daubechies-10 wavelets, learnable asymmetric hard thresholding (LAHT), and various kernel learning strategies, has been shown to effectively capture the nuances of emotional expressions in speech. Through the ablation study, we observed that the integration of LAHT and the strategic learning of kernels significantly enhanced the model's adaptability and effectiveness. Moreover, the inclusion of Bi-GRU and temporal attention layers contributed to capturing temporal dynamics and context in speech. Our comparison with other state-of-the-art methods in SER further highlights the competitiveness of SigWavNet. By outperforming these methods, SigWavNet establishes itself as a reliable solution, capable of addressing the complexities associated with SER. In conclusion, SigWavNet's ability to accurately recognize and classify emotions in speech, backed by robust empirical evidence, makes it a potential choice for various applications, ranging from human-computer interaction to mental health assessment. Future work may explore the potential of SigWavNet in more diverse and challenging real-world scenarios, as well as its adaptability to different languages and dialects.

\appendix{
\label{appendix}
This Appendix details the 1D dilated CNN with spatial attention and the Bi-GRU with temporal attention mechanisms. These components complement the learnable FDWT layer by capturing local and temporal dependencies and emphasizing key features. Their detailed formulations and implementation are provided below.

\subsubsection{1D dilated CNN and spatial attention}

The second component of our architecture is a 1-dimensional dilated CNN (1D-dilated CNN), which plays a pivotal role in our framework. To optimize computational efficiency in subsequent layers, we have strategically reduced temporal resolution in the top layers. This reduction is achieved by employing large strides in convolutional layers, along with an extensive receptive field and dilation, ensuring broad coverage of input data.

In our model, if the filter width is denoted as \(S\), then the weight tensor (Weight) has dimensions \((K, C, S)\). The standard 1D convolution layer, which can be represented mathematically as shown in eq. \eqref{eq11}, involves a direct multiplication of the filter weights with the input tensor:
\begin{equation}
\operatorname{Out}(k, q)=\sum_c \sum_{w+s=q} \operatorname{In}(c, w) * \operatorname{Weight}(k, c, s)
\label{eq11}
\end{equation}

\noindent In contrast, the 1D dilated convolution layer introduces a dilation amount \(d\), where filter weights are applied to every \(d^{th}\) element of the input tensor along its width dimension. This can be mathematically represented in eq. \eqref{eq12} as:
\begin{equation}
\operatorname{Out}(k, q)=\sum_c \sum_{w+d * s=q} \operatorname{In}(c, w) * \operatorname{Weight}(k, c, s)
\label{eq12}
\end{equation}

Dilated convolution expands the filter’s span without increasing the number of weight parameters, thereby enlarging the receptive field of the neural network without additional computational cost. This attribute is particularly advantageous for capturing long-range dependencies and contextual information in speech signals. Emotional cues in speech often span over extended periods, and dilated CNNs are adept at discerning these cues, capturing both fine-grained details and broader context effectively. Furthermore, the dilated architecture naturally accommodates variable-length sequences, which is crucial given the varying durations of emotional expressions in speech. This adaptability, combined with the network's ability to learn hierarchical features across different scales, significantly enhances the model's proficiency in extracting discriminative patterns crucial for emotion classification. Standard 1D convolution can be conceptualized as a specific case of 1D dilated convolution with a dilation factor \(d = 1\), as evident from eqs. \eqref{eq11} and \eqref{eq12}.

After each convolutional operation, we apply Instance Normalization (IN), which stabilizes training, enhances generalization across various speakers and conditions, and improves the model's robustness to scale and shift variations \cite{sefara2019effects}. IN normalizes activations along the time steps in 1D CNN, making the model less sensitive to varying recording conditions. Additionally, our model incorporates the Leaky ReLU activation function. This function addresses the "dying ReLU" problem by maintaining a non-zero gradient for negative inputs, thereby enhancing training efficiency and maintaining a mean activation closer to zero \cite{xu2015empirical} \cite{maas2013rectifier}.

To further optimize the model, we implement several dilated convolutional layers with smaller receptive fields after the initial convolutional layer, reducing the total number of parameters. Following the dilated convolutional layers, we integrate a 1D adaptation of spatial attention mechanism \cite{jetley2018learn}. This layer focuses the model's processing power on significant features by weighting different parts of the input data differently. Spatial attention dynamically recalibrates the model’s focus, directing it toward the most informative parts of the input signal. The procedure commences with the amalgamation of an input feature map, \(l\), characterized by the dimensions for batch size (\(N\)), number of channels (\(C\)), and width (\(W\)), with a global context, \(g\) as shown in eq. \eqref{eq13}. This combination is then subjected to a 1D convolution operation, utilizing a kernel size of 1, to yield a raw attention map, \(c\) (see eq. \eqref{eq14}). This convolutional step is essential for blending both local and global contextual information, laying the groundwork for attention weighting.
\begin{equation}
L_{combined} = l + g
\label{eq13}
\end{equation}
\begin{equation}
c = \text{Conv1D}_{\text{kernel\_size}=1}(L_{combined})
\label{eq14}
\end{equation}
\vspace{-3.5em}
\begin{center}
\begin{tabular}{p{4.5cm}p{4cm}}
\begin{equation}
a = \text{Softmax}(c)
\label{eq15}
\end{equation}
&
\begin{equation}
g = a \odot l
\label{eq16}
\end{equation}
\end{tabular}
\end{center}
\vspace{-2em}
\begin{equation}
G_{aggregated} = \sum_{\text{width}} g
\label{eq17}
\end{equation}

With the adoption of normalization within our mechanism, the raw attention map, \(c\), is transformed into an attention scores map, \(a\), through the softmax function as formulated in eq. \eqref{eq15}. This normalization ensures the proportional distribution of attention scores across the width, \(W\), of the input feature map. The softmax function is instrumental in prioritizing certain features by dynamically adjusting the model's focus based on the computed attention scores. The derived attention scores then modulate the input feature map, \(l\), in an element-wise manner, producing an attention-weighted feature map, \(g\) (see eq. \eqref{eq16}). This process of element-wise multiplication between the attention scores and the input features amplifies regions deemed more relevant for the task of emotion recognition. Subsequently, with normalization in effect, the aggregation of this attention-weighted feature map is executed by summing across the width (\(W\)) as shown in eq. \eqref{eq17}. This summation compresses the signal, preserving the most vital information as highlighted by the spatial attention mechanism.

This mechanism is particularly useful in scenarios where not all parts of the input signal are equally relevant for emotion recognition, such as in cases where certain speech segments carry more emotional weight than others. By employing spatial attention, our model becomes more adept at discerning subtle emotional nuances, thereby enhancing its overall predictive performance and making it more efficient and effective at identifying the correct emotional states from speech signals.

\noindent Our architecture distinctively omits pooling layers, a choice that substantively contributes to several key advantages in processing and interpreting speech signals. This decision aids in the preservation of temporal resolution, ensuring that the detailed timing of speech features is retained throughout the analysis. It also guarantees the retention of detailed information, allowing the model to access and utilize the full depth of data within the speech signal. Improved temporal localization is another significant benefit, as the model can more accurately identify and react to specific moments within the input sequence. The flexibility in handling sequence lengths is enhanced, as the model can accommodate inputs of varying durations without the need for adjustments that pooling layers might necessitate. Lastly, this architecture allows for enhanced learning from high-frequency components of the speech signal, which are crucial for distinguishing subtle emotional cues.

\subsubsection{Bi-GRU and temporal attention}
The third integral component of our architecture is the Bidirectional Gated Recurrent Units (Bi-GRU) Neural Network, as shown in Fig. \ref{fig4}, positioned after the 1D dilated CNN and the spatial attention layer. GRU operations are fundamental to this component, and their functionality is encapsulated in eqs. \eqref{eq18}, \eqref{eq19}, \eqref{eq20}, and \eqref{eq21}:
\begin{equation}
r_t = \sigma(W_{ir} x_t + b_{ir}  + W_{hr} h_{(t-1)} + b_{hr})
\label{eq18}
\end{equation}
\begin{equation}
z_t = \sigma(W_{iz} + b_{iz} + W_{hz} h_{(t-1)} + b_{hz})
\label{eq19}
\end{equation}
\begin{equation}
n_t = \tanh(W_{in} x_t + b_{in} + r_t * (W_{hn} h_{(t-1)} + b_{hn}))
\label{eq20}
\end{equation}
\begin{equation}
h_t = (1 - z_t) * n_t + z_t * h_{(t-1)}
\label{eq21}
\end{equation}

\noindent where, \(h_t\) represents the hidden state at time \(t\), \(x_t\) is the input at time \(t\), and \(h_{(t-1)}\) is the hidden state from the previous time step. The terms \(r_t\), \(z_t\), and \(n_t\) correspond to the reset, update, and new gates, respectively. The sigmoid function is denoted by \(\sigma\), and the Hadamard product is represented by \( *\) \cite{styan1973hadamard}. In a multilayer GRU, like ours with \(n\) layers, the input \(x^{(l)}_t\) for layer \(l\) (where \(l \geq 2\)) is the hidden state from the preceding layer \(h^{(l-1)}_t\), modified by dropout \(\delta^{(l-1)}_t\). Each \(\delta^{(l-1)}_t\) is a Bernoulli random variable that equals \(0\) with a dropout probability \cite{chung2014empirical}.

The Bi-GRU operates by processing data through two unidirectional GRUs moving in opposite directions, one forward from the start of the data sequence and the other backward from the end. This structure allows the network to incorporate both past and future information into the current state. The Bi-GRU is formulated in eqs. \eqref{eq22}, \eqref{eq23}, and \eqref{eq24} as:
\begin{equation}
\overrightarrow{h_t}=G R U_{fwd}\left(x_t, \overrightarrow{h_{t-1}}\right)
\label{eq22}
\end{equation}
\begin{equation}
\overleftarrow{h_t}=G R U_{bwd}\left(x_t, \overleftarrow{h_{t+1}}\right)
\label{eq23}
\end{equation}
\begin{equation}
h_t=\overrightarrow{h_t} \oplus \overleftarrow{h_t}
\label{eq24}
\end{equation}

\noindent In these equations, \(\overrightarrow{h_t}\) and \(\overleftarrow{h_t}\) represent the states of the forward and backward GRUs, respectively, while \(\oplus\) denotes the concatenation of two vectors.

Following the incorporation of Bi-GRU into our architecture, a Temporal Attention layer is adeptly positioned to further refine the model's capacity for emotion recognition. This layer leverages the comprehensive temporal dynamics captured by the Bi-GRU, applying focused attention to the most salient features across the time sequence. It initiates its process by subjecting the sequence of hidden states, \(H \in \mathbb{R}^{N \times T \times D}\) where \(N\) is the batch size, \(T\) denotes the time steps, and \(D\) represents the hidden state dimensionality, to a linear transformation via a fully connected layer \(\text{FC1}\), aiming to prepare the states for relevance evaluation against the sequence's final hidden state \(h_t\), encapsulating the aggregate contextual information. This step is represented as in eq. \eqref{eq25}:
\begin{equation}
    Score = \text{FC1}(H) \cdot h_t^T
    \label{eq25}
\end{equation}

\noindent where \(Score\) encapsulates the relevance scores for each time step, and \(h_t = H[:, -1, :]\) is extracted as the last hidden state from \(H\). Following score computation, a softmax function is applied to normalize these scores across the time steps, yielding attention weights, \(A\), which articulate the relative importance of each time step's features (see eq. \eqref{eq26}):
\begin{equation}
    A = \text{Softmax}(Score)
    \label{eq26}
\end{equation}
    
Utilizing these attention weights, a context vector \(C\) is formulated through a weighted summation of the hidden states as shown in eq. \eqref{eq27}, thereby distilling the sequence's most pivotal information:
\begin{equation}
    C = \sum_{t=1}^{T} A_t \cdot H_t
    \label{eq27}
\end{equation}

In the final step, this context vector is concatenated with the last hidden state, \(h_t\), and the resultant vector undergoes another linear transformation through \(\text{FC2}\) and is passed through a \(\tanh\) activation function to produce the attention vector \(V\) (see eq. \eqref{eq28}), which encapsulates the focused interpretation of the sequence:
\begin{equation}
    V = \tanh(\text{FC2}([C; h_t]))
    \label{eq28}
\end{equation}

By assessing the importance of each time step's features in relation to the overall sequence, the Temporal Attention mechanism dynamically weights these features to emphasize those most indicative of emotional states. This process not only enhances the model's sensitivity to crucial emotional cues dispersed throughout the speech signal but also facilitates a more nuanced aggregation of temporal information.

The output from each frequency band, after temporal attention, manifests as an attention vector. This vector encapsulates the quintessential information crucial for emotion recognition, derived from the respective frequency band it represents. With the FDWT providing multiple decomposition levels, each correlating to a distinct frequency band, we acquire a suite of attention vectors—one for each high-frequency representation plus an additional one for the low-frequency representation.}

\bibliographystyle{IEEEtran}
\bibliography{references}

\begin{thebibliography}{10}
\providecommand{\url}[1]{#1}
\csname url@samestyle\endcsname
\providecommand{\newblock}{\relax}
\providecommand{\bibinfo}[2]{#2}
\providecommand{\BIBentrySTDinterwordspacing}{\spaceskip=0pt\relax}
\providecommand{\BIBentryALTinterwordstretchfactor}{4}
\providecommand{\BIBentryALTinterwordspacing}{\spaceskip=\fontdimen2\font plus
\BIBentryALTinterwordstretchfactor\fontdimen3\font minus \fontdimen4\font\relax}
\providecommand{\BIBforeignlanguage}[2]{{%
\expandafter\ifx\csname l@#1\endcsname\relax
\typeout{** WARNING: IEEEtran.bst: No hyphenation pattern has been}%
\typeout{** loaded for the language `#1'. Using the pattern for}%
\typeout{** the default language instead.}%
\else
\language=\csname l@#1\endcsname
\fi
#2}}
\providecommand{\BIBdecl}{\relax}
\BIBdecl

\bibitem{nfissi2024unlocking}
A.~Nfissi, W.~Bouachir, N.~Bouguila, and B.~Mishara, ``Unlocking the emotional states of high-risk suicide callers through speech analysis,'' in \emph{2024 IEEE 18th International Conference on Semantic Computing (ICSC)}.\hskip 1em plus 0.5em minus 0.4em\relax IEEE Computer Society, 2024, pp. 33--40.

\bibitem{cdcp_2022}
``Suicide prevention (2024),'' centers for Disease Control and Prevention. Available at: https://www.cdc.gov/suicide/ (Accessed: February 10, 2024).

\bibitem{9544671}
Y.~Bhangdia, R.~Bhansali, N.~Chaudhari, D.~Chandnani, and M.~L. Dhore, ``Speech emotion recognition and sentiment analysis based therapist bot,'' in \emph{2021 Third International Conference on Inventive Research in Computing Applications (ICIRCA)}, 2021, pp. 96--101.

\bibitem{nugroho2022development}
H.~Nugroho and R.~N.~N. Fuadiyah, ``Development of speech emotion recognition system based on discrete wavelet transform (dwt) and voice segmentation,'' \emph{International Journal on Electrical Engineering and Informatics}, vol.~14, no.~3, pp. 593--607, 2022.

\bibitem{el2011survey}
M.~El~Ayadi, M.~S. Kamel, and F.~Karray, ``Survey on speech emotion recognition: Features, classification schemes, and databases,'' \emph{Pattern recognition}, vol.~44, no.~3, pp. 572--587, 2011.

\bibitem{wang2020missing}
Q.~Wang, G.~Michau, and O.~Fink, ``Missing-class-robust domain adaptation by unilateral alignment,'' \emph{IEEE Transactions on Industrial Electronics}, vol.~68, no.~1, pp. 663--671, 2020.

\bibitem{li2019deep}
Z.~Li, Y.~Wang, and K.~Wang, ``A deep learning driven method for fault classification and degradation assessment in mechanical equipment,'' \emph{Computers in industry}, vol. 104, pp. 1--10, 2019.

\bibitem{fink2020potential}
O.~Fink, Q.~Wang, M.~Svensen, P.~Dersin, W.-J. Lee, and M.~Ducoffe, ``Potential, challenges and future directions for deep learning in prognostics and health management applications,'' \emph{Engineering Applications of Artificial Intelligence}, vol.~92, p. 103678, 2020.

\bibitem{michau2021unsupervised}
G.~Michau and O.~Fink, ``Unsupervised transfer learning for anomaly detection: Application to complementary operating condition transfer,'' \emph{Knowledge-Based Systems}, vol. 216, p. 106816, 2021.

\bibitem{fahad2021survey}
M.~S. Fahad, A.~Ranjan, J.~Yadav, and A.~Deepak, ``A survey of speech emotion recognition in natural environment,'' \emph{Digital signal processing}, vol. 110, p. 102951, 2021.

\bibitem{huang2022research}
L.~Huang and X.~Shen, ``Research on speech emotion recognition based on the fractional fourier transform,'' \emph{Electronics}, vol.~11, no.~20, p. 3393, 2022.

\bibitem{kiranyaz20211d}
S.~Kiranyaz, O.~Avci, O.~Abdeljaber, T.~Ince, M.~Gabbouj, and D.~J. Inman, ``1d convolutional neural networks and applications: A survey,'' \emph{Mechanical systems and signal processing}, vol. 151, p. 107398, 2021.

\bibitem{zhang2018deep}
W.~Zhang, C.~Li, G.~Peng, Y.~Chen, and Z.~Zhang, ``A deep convolutional neural network with new training methods for bearing fault diagnosis under noisy environment and different working load,'' \emph{Mechanical systems and signal processing}, vol. 100, pp. 439--453, 2018.

\bibitem{papyan2017convolutional}
V.~Papyan, Y.~Romano, and M.~Elad, ``Convolutional neural networks analyzed via convolutional sparse coding,'' \emph{The Journal of Machine Learning Research}, vol.~18, no.~1, pp. 2887--2938, 2017.

\bibitem{uteuliyeva2020fourier}
M.~Uteuliyeva, A.~Zhumekenov, R.~Takhanov, Z.~Assylbekov, A.~J. Castro, and O.~Kabdolov, ``Fourier neural networks: A comparative study,'' \emph{Intelligent Data Analysis}, vol.~24, no.~5, pp. 1107--1120, 2020.

\bibitem{li2021waveletkernelnet}
T.~Li, Z.~Zhao, C.~Sun, L.~Cheng, X.~Chen, R.~Yan, and R.~X. Gao, ``Waveletkernelnet: An interpretable deep neural network for industrial intelligent diagnosis,'' \emph{IEEE Transactions on Systems, Man, and Cybernetics: Systems}, vol.~52, no.~4, pp. 2302--2312, 2021.

\bibitem{liu2019multi}
P.~Liu, H.~Zhang, W.~Lian, and W.~Zuo, ``Multi-level wavelet convolutional neural networks,'' \emph{IEEE Access}, vol.~7, pp. 74\,973--74\,985, 2019.

\bibitem{khalil2020end}
M.~Khalil, A.~Adib \emph{et~al.}, ``An end-to-end multi-level wavelet convolutional neural networks for heart diseases diagnosis,'' \emph{Neurocomputing}, vol. 417, pp. 187--201, 2020.

\bibitem{gao2017speech}
Y.~Gao, B.~Li, N.~Wang, and T.~Zhu, ``Speech emotion recognition using local and global features,'' in \emph{Brain Informatics: International Conference, BI 2017, Beijing, China, November 16-18, 2017, Proceedings}.\hskip 1em plus 0.5em minus 0.4em\relax Springer, 2017, pp. 3--13.

\bibitem{eyben2016open}
F.~Eyben, F.~Weninger, M.~W{\"o}llmer, and B.~Shuller, ``Open-source media interpretation by large feature-space extraction,'' \emph{TU Munchen, MMK}, 2016.

\bibitem{zhang2017speech}
S.~Zhang, S.~Zhang, T.~Huang, and W.~Gao, ``Speech emotion recognition using deep convolutional neural network and discriminant temporal pyramid matching,'' \emph{IEEE Transactions on Multimedia}, vol.~20, no.~6, pp. 1576--1590, 2017.

\bibitem{zhao2019speech}
J.~Zhao, X.~Mao, and L.~Chen, ``Speech emotion recognition using deep 1d \& 2d cnn lstm networks,'' \emph{Biomedical signal processing and control}, vol.~47, pp. 312--323, 2019.

\bibitem{schuller2003hidden}
B.~Schuller, G.~Rigoll, and M.~Lang, ``Hidden markov model-based speech emotion recognition,'' in \emph{2003 IEEE International Conference on Acoustics, Speech, and Signal Processing, 2003. Proceedings.(ICASSP'03).}, vol.~2.\hskip 1em plus 0.5em minus 0.4em\relax Ieee, 2003, pp. II--1.

\bibitem{rolnick2017deep}
D.~Rolnick, A.~Veit, S.~Belongie, and N.~Shavit, ``Deep learning is robust to massive label noise,'' \emph{arXiv preprint arXiv:1705.10694}, 2017.

\bibitem{mustaqeem2019cnn}
Mustaqeem and S.~Kwon, ``A cnn-assisted enhanced audio signal processing for speech emotion recognition,'' \emph{Sensors}, vol.~20, no.~1, p. 183, 2019.

\bibitem{mallat2010recursive}
S.~Mallat, ``Recursive interferometric representation,'' in \emph{Proc. of EUSICO conference, Danemark}, vol.~3, 2010.

\bibitem{mallat2012group}
------, ``Group invariant scattering,'' \emph{Communications on Pure and Applied Mathematics}, vol.~65, no.~10, pp. 1331--1398, 2012.

\bibitem{anden2014deep}
J.~And{\'e}n and S.~Mallat, ``Deep scattering spectrum,'' \emph{IEEE Transactions on Signal Processing}, vol.~62, no.~16, pp. 4114--4128, 2014.

\bibitem{anden2019joint}
J.~And{\'e}n, V.~Lostanlen, and S.~Mallat, ``Joint time--frequency scattering,'' \emph{IEEE Transactions on Signal Processing}, vol.~67, no.~14, pp. 3704--3718, 2019.

\bibitem{ghezaiel2021hybrid}
W.~Ghezaiel, L.~Brun, and O.~L{\'e}zoray, ``Hybrid network for end-to-end text-independent speaker identification,'' in \emph{2020 25th International conference on pattern recognition (ICPR)}.\hskip 1em plus 0.5em minus 0.4em\relax IEEE, 2021, pp. 2352--2359.

\bibitem{silva2009discriminative}
J.~Silva and S.~S. Narayanan, ``Discriminative wavelet packet filter bank selection for pattern recognition,'' \emph{IEEE Transactions on Signal Processing}, vol.~57, no.~5, pp. 1796--1810, 2009.

\bibitem{rao2018discrete}
K.~D. Rao, M.~Swamy, K.~D. Rao, and M.~Swamy, ``Discrete wavelet transforms,'' \emph{Digital Signal Processing: Theory and Practice}, pp. 619--691, 2018.

\bibitem{daubechies1990wavelet}
I.~Daubechies, ``The wavelet transform, time-frequency localization and signal analysis,'' \emph{IEEE transactions on information theory}, vol.~36, no.~5, pp. 961--1005, 1990.

\bibitem{zao2014time}
L.~Z{\~a}o, D.~Cavalcante, and R.~Coelho, ``Time-frequency feature and ams-gmm mask for acoustic emotion classification,'' \emph{IEEE signal processing letters}, vol.~21, no.~5, pp. 620--624, 2014.

\bibitem{muthusamy2015particle}
H.~Muthusamy, K.~Polat, and S.~Yaacob, ``Particle swarm optimization based feature enhancement and feature selection for improved emotion recognition in speech and glottal signals,'' \emph{PloS one}, vol.~10, no.~3, p. e0120344, 2015.

\bibitem{zheng2018effectiveness}
B.~S. Zheng, W.~Khairunizam, S.~Murugappan~Murugappan, Z.~Razlan, I.~Zunaidi, and C.~Yean, ``Effectiveness of tuned q-factor wavelet transform in emotion recognition among left-brain damaged stroke patients,'' \emph{Int J Simul Syst Sci Technol}, vol.~19, no.~3, p.~2, 2018.

\bibitem{cowie2001emotion}
R.~Cowie, E.~Douglas-Cowie, N.~Tsapatsoulis, G.~Votsis, S.~Kollias, W.~Fellenz, and J.~G. Taylor, ``Emotion recognition in human-computer interaction,'' \emph{IEEE Signal processing magazine}, vol.~18, no.~1, pp. 32--80, 2001.

\bibitem{mansoorizadeh2007speech}
M.~Mansoorizadeh and N.~M. Charkari, ``Speech emotion recognition: Comparison of speech segmentation approaches,'' \emph{Proceedings of IKT, Mashad, Iran}, 2007.

\bibitem{steinbuch2005wavelet}
M.~Steinbuch and M.~Van~de Molengraft, ``Wavelet theory and applications: a literature study,'' \emph{Eindhoven: Eindhoven University Technology Department of Mechanical Engineering Control System Group}, 2005.

\bibitem{palo2018wavelet}
H.~K. Palo and M.~N. Mohanty, ``Wavelet based feature combination for recognition of emotions,'' \emph{Ain shams engineering journal}, vol.~9, no.~4, pp. 1799--1806, 2018.

\bibitem{abdel2020egyptian}
L.~Abdel-Hamid, ``Egyptian arabic speech emotion recognition using prosodic, spectral and wavelet features,'' \emph{Speech Communication}, vol. 122, pp. 19--30, 2020.

\bibitem{wang2020wavelet}
K.~Wang, G.~Su, L.~Liu, and S.~Wang, ``Wavelet packet analysis for speaker-independent emotion recognition,'' \emph{Neurocomputing}, vol. 398, pp. 257--264, 2020.

\bibitem{kishore2013emotion}
K.~K. Kishore and P.~K. Satish, ``Emotion recognition in speech using mfcc and wavelet features,'' in \emph{2013 3rd IEEE International Advance Computing Conference (IACC)}.\hskip 1em plus 0.5em minus 0.4em\relax IEEE, 2013, pp. 842--847.

\bibitem{huang2019feature}
Y.~Huang, K.~Tian, A.~Wu, and G.~Zhang, ``Feature fusion methods research based on deep belief networks for speech emotion recognition under noise condition,'' \emph{Journal of ambient intelligence and humanized computing}, vol.~10, pp. 1787--1798, 2019.

\bibitem{vasquez2015emotion}
J.~C. V{\'a}squez-Correa, N.~Garc{\'\i}a, J.~R. Orozco-Arroyave, J.~D. Arias-Londo{\~n}o, J.~F. Vargas-Bonilla, and E.~N{\"o}th, ``Emotion recognition from speech under environmental noise conditions using wavelet decomposition,'' in \emph{2015 International Carnahan Conference on Security Technology (ICCST)}.\hskip 1em plus 0.5em minus 0.4em\relax IEEE, 2015, pp. 247--252.

\bibitem{busch2007heisenberg}
P.~Busch, T.~Heinonen, and P.~Lahti, ``Heisenberg's uncertainty principle,'' \emph{Physics reports}, vol. 452, no.~6, pp. 155--176, 2007.

\bibitem{mallat2008wavelet}
\BIBentryALTinterwordspacing
S.~Mallat, \emph{A Wavelet Tour of Signal Processing: The Sparse Way}.\hskip 1em plus 0.5em minus 0.4em\relax Elsevier Science, 2008. [Online]. Available: \url{https://books.google.ca/books?id=5qzeLJljuLoC}
\BIBentrySTDinterwordspacing

\bibitem{croisier1976perfect}
A.~Croisier, ``Perfect channel splitting by use of interpolation/decimation/tree decomposition techniques,'' in \emph{Proc. Int. Symp. on Info., Circuits and Systems,(Patras, Greece)}, 1976.

\bibitem{lin2013network}
M.~Lin, Q.~Chen, and S.~Yan, ``Network in network,'' \emph{arXiv preprint arXiv:1312.4400}, 2013.

\bibitem{singh2022systematic}
Y.~B. Singh and S.~Goel, ``A systematic literature review of speech emotion recognition approaches,'' \emph{Neurocomputing}, 2022.

\bibitem{busso2008iemocap}
C.~Busso, M.~Bulut, C.-C. Lee, A.~Kazemzadeh, E.~Mower, S.~Kim, J.~N. Chang, S.~Lee, and S.~S. Narayanan, ``Iemocap: Interactive emotional dyadic motion capture database,'' \emph{Language resources and evaluation}, vol.~42, no.~4, pp. 335--359, 2008.

\bibitem{7178872}
Q.~Jin, C.~Li, S.~Chen, and H.~Wu, ``Speech emotion recognition with acoustic and lexical features,'' in \emph{2015 IEEE International Conference on Acoustics, Speech and Signal Processing (ICASSP)}, 2015, pp. 4749--4753.

\bibitem{kim2013deep}
Y.~Kim, H.~Lee, and E.~M. Provost, ``Deep learning for robust feature generation in audiovisual emotion recognition,'' in \emph{2013 IEEE international conference on acoustics, speech and signal processing}.\hskip 1em plus 0.5em minus 0.4em\relax IEEE, 2013, pp. 3687--3691.

\bibitem{burkhardt2005database}
F.~Burkhardt, A.~Paeschke, M.~Rolfes, W.~F. Sendlmeier, B.~Weiss \emph{et~al.}, ``A database of german emotional speech.'' in \emph{Interspeech}, vol.~5, 2005, pp. 1517--1520.

\bibitem{aoyama1954study}
H.~Aoyama, ``A study of stratified random sampling,'' \emph{Ann. Inst. Stat. Math}, vol.~6, no.~1, pp. 1--36, 1954.

\bibitem{li2018massively}
L.~Li, K.~Jamieson, A.~Rostamizadeh, E.~Gonina, M.~Hardt, B.~Recht, and A.~Talwalkar, ``Massively parallel hyperparameter tuning,'' 2018.

\bibitem{li2020system}
L.~Li, K.~Jamieson, A.~Rostamizadeh, E.~Gonina, J.~Ben-Tzur, M.~Hardt, B.~Recht, and A.~Talwalkar, ``A system for massively parallel hyperparameter tuning,'' \emph{Proceedings of Machine Learning and Systems}, vol.~2, pp. 230--246, 2020.

\bibitem{lin2017focal}
T.-Y. Lin, P.~Goyal, R.~Girshick, K.~He, and P.~Doll{\'a}r, ``Focal loss for dense object detection,'' in \emph{Proceedings of the IEEE international conference on computer vision}, 2017, pp. 2980--2988.

\bibitem{https://doi.org/10.48550/arxiv.1412.6980}
\BIBentryALTinterwordspacing
D.~P. Kingma and J.~Ba, ``Adam: A method for stochastic optimization,'' 2014. [Online]. Available: \url{https://arxiv.org/abs/1412.6980}
\BIBentrySTDinterwordspacing

\bibitem{aftab2022light}
A.~Aftab, A.~Morsali, S.~Ghaemmaghami, and B.~Champagne, ``Light-sernet: A lightweight fully convolutional neural network for speech emotion recognition,'' in \emph{ICASSP 2022-2022 IEEE International Conference on Acoustics, Speech and Signal Processing (ICASSP)}.\hskip 1em plus 0.5em minus 0.4em\relax IEEE, 2022, pp. 6912--6916.

\bibitem{issa2020speech}
D.~Issa, M.~F. Demirci, and A.~Yazici, ``Speech emotion recognition with deep convolutional neural networks,'' \emph{Biomedical Signal Processing and Control}, vol.~59, p. 101894, 2020.

\bibitem{chen20183}
M.~Chen, X.~He, J.~Yang, and H.~Zhang, ``3-d convolutional recurrent neural networks with attention model for speech emotion recognition,'' \emph{IEEE Signal Processing Letters}, vol.~25, no.~10, pp. 1440--1444, 2018.

\bibitem{meng2019speech}
H.~Meng, T.~Yan, F.~Yuan, and H.~Wei, ``Speech emotion recognition from 3d log-mel spectrograms with deep learning network,'' \emph{IEEE access}, vol.~7, pp. 125\,868--125\,881, 2019.

\bibitem{li2019improved}
Y.~Li, T.~Zhao, and T.~Kawahara, ``Improved end-to-end speech emotion recognition using self attention mechanism and multitask learning.'' in \emph{Interspeech}, 2019, pp. 2803--2807.

\bibitem{singh2021deep}
P.~Singh, G.~Saha, and M.~Sahidullah, ``Deep scattering network for speech emotion recognition,'' in \emph{2021 29th European Signal Processing Conference (EUSIPCO)}.\hskip 1em plus 0.5em minus 0.4em\relax IEEE, 2021, pp. 131--135.

\bibitem{meng2021speech}
H.~Meng, T.~Yan, H.~Wei, and X.~Ji, ``Speech emotion recognition using wavelet packet reconstruction with attention-based deep recurrent neutral networks,'' \emph{Bulletin of the Polish Academy of Sciences. Technical Sciences}, vol.~69, no.~1, 2021.

\bibitem{Iqbal2020MFCCAM}
M.~Z. Iqbal, ``Mfcc and machine learning based speech emotion recognition over tess and iemocap datasets,'' 2020.

\bibitem{parlak2014cross}
C.~Parlak, B.~Diri, and F.~G{\"u}rgen, ``A cross-corpus experiment in speech emotion recognition.'' in \emph{SLAM@ INTERSPEECH}, 2014, pp. 58--61.

\bibitem{pham2021emotion}
M.~H. Pham, F.~M. Noori, and J.~Torresen, ``Emotion recognition using speech data with convolutional neural network,'' in \emph{2021 IEEE 2nd International Conference on Signal, Control and Communication (SCC)}.\hskip 1em plus 0.5em minus 0.4em\relax IEEE, 2021, pp. 182--187.

\bibitem{ancilin2021improved}
J.~Ancilin and A.~Milton, ``Improved speech emotion recognition with mel frequency magnitude coefficient,'' \emph{Applied Acoustics}, vol. 179, p. 108046, 2021.

\bibitem{van2023speech}
B.~E. Van~Zwol, M.~A. Langezaal, L.~Arts, A.~Gatt, and E.~L. Van Den~Broek, ``Speech emotion recognition using deep convolutional neural networks improved by the fast continuous wavelet transform,'' in \emph{Workshop Proceedings of the 19th International Conference on Intelligent Environments (IE2023)}.\hskip 1em plus 0.5em minus 0.4em\relax IOS Press, 2023, pp. 63--72.

\bibitem{tuncer2021automated}
T.~Tuncer, S.~Dogan, and U.~R. Acharya, ``Automated accurate speech emotion recognition system using twine shuffle pattern and iterative neighborhood component analysis techniques,'' \emph{Knowledge-Based Systems}, vol. 211, p. 106547, 2021.

\bibitem{palo2023amalgamation}
H.~K. Palo, S.~Subudhiray, and N.~Das, ``The amalgamation of wavelet packet information gain entropy tuned source and system parameters for improved speech emotion recognition,'' \emph{Speech Communication}, vol. 149, pp. 11--28, 2023.

\bibitem{liu2022speech}
Y.~Liu and Z.~Kexin, ``Speech emotion recognition system based on wavelet transform and multi-task learning,'' in \emph{2022 7th International Conference on Intelligent Informatics and Biomedical Science (ICIIBMS)}, vol.~7.\hskip 1em plus 0.5em minus 0.4em\relax IEEE, 2022, pp. 141--149.

\bibitem{sefara2019effects}
T.~J. Sefara, ``The effects of normalisation methods on speech emotion recognition,'' in \emph{2019 International multidisciplinary information technology and engineering conference (IMITEC)}.\hskip 1em plus 0.5em minus 0.4em\relax IEEE, 2019, pp. 1--8.

\bibitem{xu2015empirical}
B.~Xu, N.~Wang, T.~Chen, and M.~Li, ``Empirical evaluation of rectified activations in convolutional network,'' \emph{arXiv preprint arXiv:1505.00853}, 2015.

\bibitem{maas2013rectifier}
A.~L. Maas, A.~Y. Hannun, A.~Y. Ng \emph{et~al.}, ``Rectifier nonlinearities improve neural network acoustic models,'' in \emph{Proc. icml}, vol.~30, no.~1.\hskip 1em plus 0.5em minus 0.4em\relax Citeseer, 2013, p.~3.

\bibitem{jetley2018learn}
S.~Jetley, N.~A. Lord, N.~Lee, and P.~H. Torr, ``Learn to pay attention,'' \emph{arXiv preprint arXiv:1804.02391}, 2018.

\bibitem{styan1973hadamard}
G.~P. Styan, ``Hadamard products and multivariate statistical analysis,'' \emph{Linear algebra and its applications}, vol.~6, pp. 217--240, 1973.

\bibitem{chung2014empirical}
J.~Chung, C.~Gulcehre, K.~Cho, and Y.~Bengio, ``Empirical evaluation of gated recurrent neural networks on sequence modeling,'' \emph{arXiv preprint arXiv:1412.3555}, 2014.

\end{thebibliography}

\vspace{1em}
\section{Biography}
\vspace{-4.5em}
\begin{IEEEbiography}[{\includegraphics[width=1in,height=1.25in,clip,keepaspectratio]{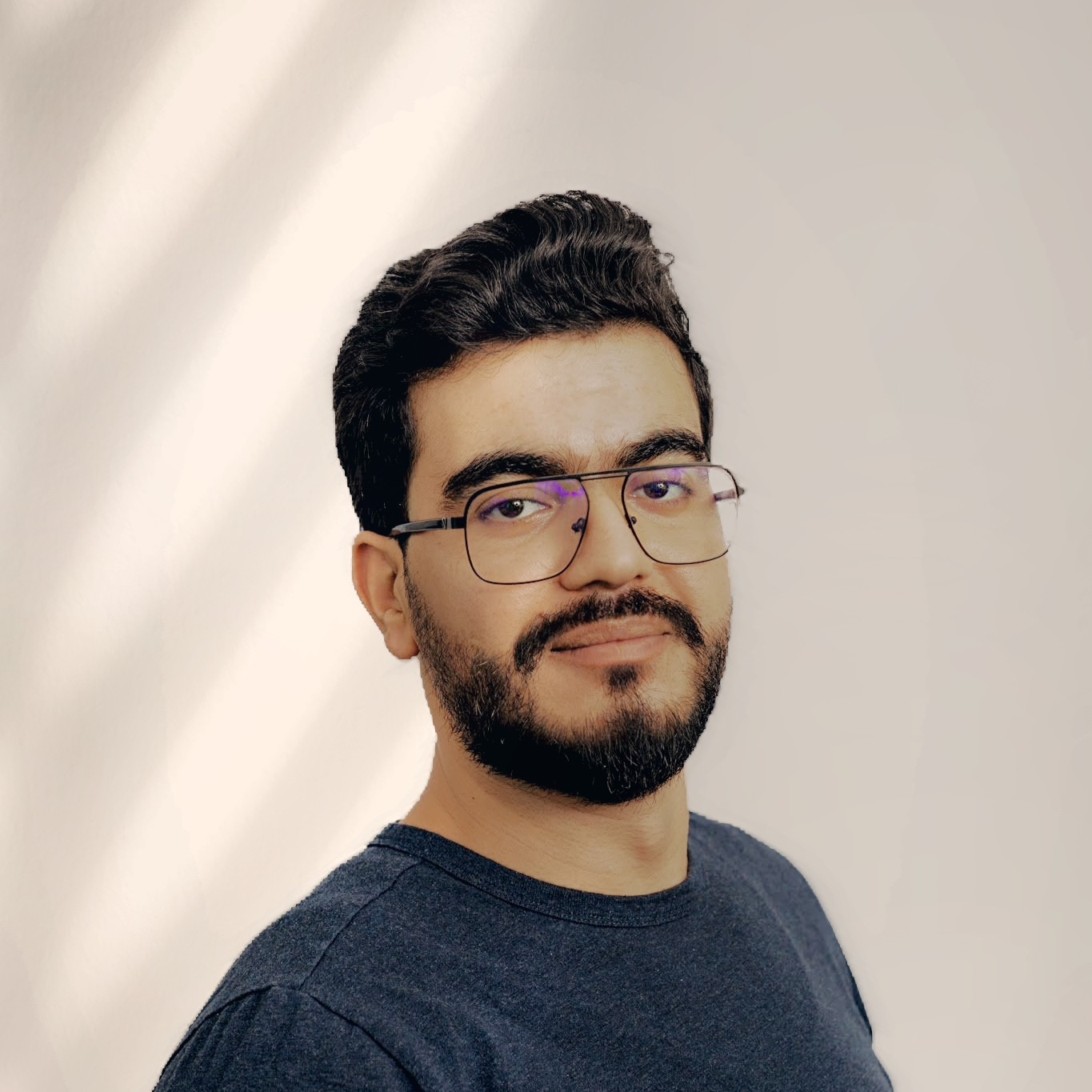}}]{Alaa Nfissi}
received an engineering degree in computer science and data science from the Private High School of Engineering and Technologies (Tunisia) in 2020. He then earned his MSc degree in advanced engineering, internet of things, and data processing, from the Polytechnic School of Tunisia in 2021. Currently, he is pursuing a Ph.D. at Concordia University, Montreal, Quebec, Canada. His research in artificial intelligence includes speech and emotion analysis, signal processing, and pattern recognition.
\end{IEEEbiography}
\vspace{-4.5em}
\begin{IEEEbiography}[{\includegraphics[width=1in,height=1.25in,clip,keepaspectratio]{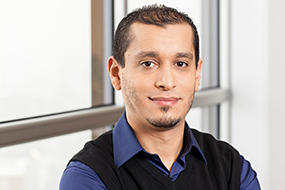}}]{Wassim Bouachir} is a professor of computer science at the University of Quebec (TELUQ) and the Canada Research Chair (CRC) in Artificial Intelligence for Suicide Prevention. He holds a Ph.D. degree in computer engineering from Polytechnique Montreal and a M.Sc. in computer science from the University of Moncton. His research interests include computer vision, signal processing, and machine learning. His research works aim to develop AI-based systems for several application areas, such as physical and mental health, security, and environment applications.
\end{IEEEbiography}
\vspace{-4.5em}
\begin{IEEEbiography}[{\includegraphics[width=1in,height=1.25in,clip,keepaspectratio]{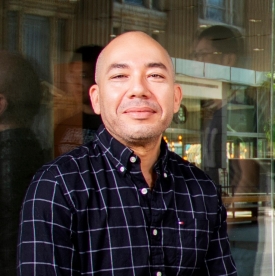}}]{Nizar Bouguila}
received the engineering degree from the University of Tunis, Tunis, Tunisia, in 2000, and the M.Sc. and Ph.D. degrees in computer science from Sherbrooke University, Sherbrooke, QC, Canada, in 2002 and 2006, respectively. He is currently a Professor with the Concordia Institute for Information Systems Engineering (CIISE) at Concordia University, Montreal, Quebec, Canada. His research interests include image processing, machine learning, data mining, computer vision, pattern recognition, smart buildings, and energy. Prof. Bouguila received the best Ph.D. thesis Award in Engineering and Natural Sciences from Sherbrooke University in 2007. He was awarded the prestigious Prix d’excellence de l’association des doyens des etudes superieures au Quebec (best Ph.D. thesis Award in Engineering and Natural Sciences in Quebec), and was a runner-up for the prestigious NSERC doctoral prize. He was the holder of a Concordia University Research Chair Tier 2 from 2014 to 2019 and was named Concordia University Research Fellow in 2020. He is currently the holder of a Concordia University Research Chair Tier 1 in Applied Artificial Intelligence. He is the author or co-author of over 500 publications in several prestigious journals and conferences. He is a regular reviewer for many international journals and serves as associate editor for several journals such as IEEE Transactions on Neural Networks and Learning Systems, Pattern Recognition, and Engineering Applications of Artificial Intelligence.
\end{IEEEbiography}
\vspace{-3.5em}
\begin{IEEEbiography}[{\includegraphics[width=1in,height=1.25in,clip,keepaspectratio]{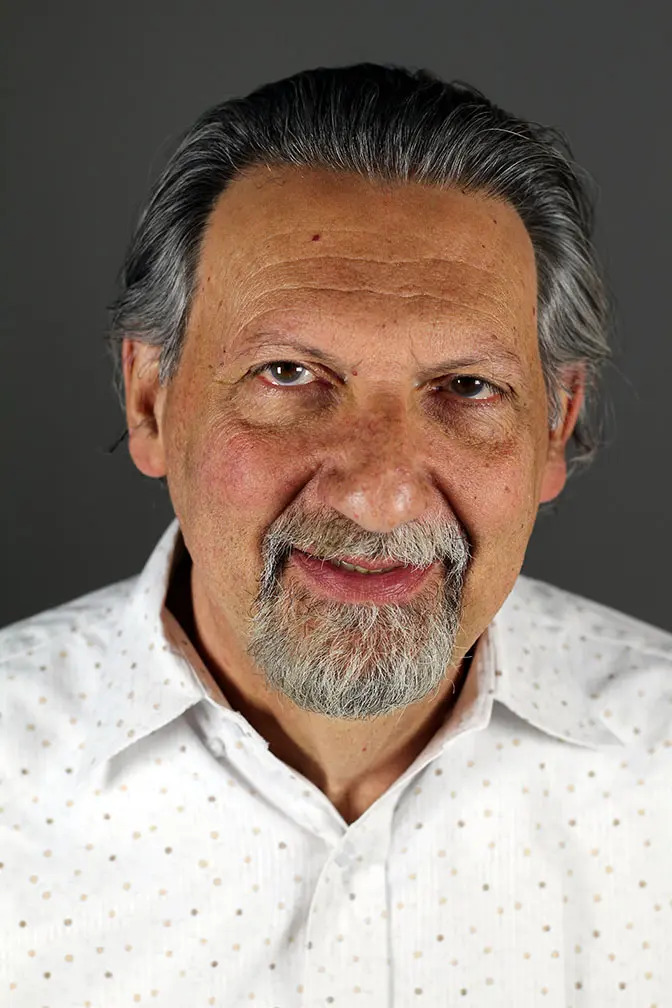}}]{Brian L. Mishara} 
is Director of the Centre for Research and Intervention on Suicide, Ethical Issues and End-of-Life Practices (CRISE), and Psychology Professor at the Université du Québec à Montréal (UQAM) in Montreal, Canada. Professor Mishara has published more than 200 book chapters and articles in peer-reviewed journals, 7 books in English and 5 in French. He was President of the International Association for Suicide Prevention (IASP), President of Canadian Association for Suicide Prevention (CASP), and Vice-Chairperson, Trustees of Befrienders Worldwide. He consults in the development of local and national suicide prevention initiatives worldwide and conducts suicide prevention training internationally.
\end{IEEEbiography}

\newpage

\end{document}